\documentclass[%
    twocolumn,
superscriptaddress,
preprintnumbers,
nofootinbib,
 amsmath,amssymb,
 aps,
 prd,
 floatfix,
]{revtex4-2}

\usepackage{graphicx}% Include figure files
\usepackage{duckuments} % DEBUG: removeme
\usepackage{dcolumn}% Align table columns on decimal point
\usepackage{bm}% bold math
\usepackage{braket}
\usepackage{mathtools} % ensuremath etc
\usepackage{enumitem} % for bullet dedentation

\usepackage[pdftex]{xcolor}
\usepackage{hyperref}% add hypertext capabilities
\hypersetup{
    colorlinks=true,     % false: boxed links; true: colored links
    linkcolor=blue,      % color of internal links
    citecolor=blue,      % color of links to bibliography
    filecolor=blue,      % color of file links
    urlcolor=blue        % color of external links
}

\usepackage{cleveref}
\crefname{equation}{Eq.}{Eqs.}
\crefname{figure}{Fig.}{Figs.}
\crefname{table}{Tab.}{Tabs.}
\crefname{section}{Sec.}{Secs.}
\crefname{appendix}{App.}{Apps.}
\Crefname{table}{Table}{Tables}
\Crefname{figure}{Figure}{Figures}

\newcommand{\aset}[1]{\{#1\}}
\newcommand{\RRvec}[0]{\xi}
\newcommand{\RRmat}[0]{E}
\newcommand{\Pm}[1]{\mathcal{P}^{(#1)}}
\newcommand{\HEq}[0]{\rightarrow}
\newcommand{\Span}{\mathrm{span}}
\newcommand{\herm}{\mathfrak{h}}

\definecolor{red}{rgb}{1.0, 0, 0}

\begin{document}

\preprint{FERMILAB-PUB-25-0131-T,MIT-CTP/5849}

% \title{Coincidences in Krylov Space\\(or, Rayleigh-Ritz is all you need)}
\title{Filtered Rayleigh-Ritz is all you need}

\author{Ryan Abbott}
 \affiliation{Center for Theoretical Physics, Massachusetts Institute of Technology, Cambridge, MA 02139, USA}
\author{Daniel C. Hackett}
 \affiliation{Fermi National Accelerator Laboratory, Batavia, IL 60510, U.S.A.}
\author{George T.\ Fleming}
 \affiliation{Fermi National Accelerator Laboratory, Batavia, IL 60510, U.S.A.}
\author{Dimitra A. Pefkou}
 \affiliation{Department of Physics, University of California, Berkeley, CA 94720, USA}
 \affiliation{Nuclear Science Division, Lawrence Berkeley National Laboratory, Berkeley, CA 94720, USA}
\author{Michael L. Wagman}
 \affiliation{Fermi National Accelerator Laboratory, Batavia, IL 60510, U.S.A.}

\begin{abstract}
    Recent work~\cite{Wagman:2024rid,Hackett:2024xnx,Hackett:2024nbe} has shown that the (block) Lanczos algorithm can be used to extract approximate energy spectra and matrix elements from (matrices of) correlation functions in quantum field theory, and identified exact coincidences between Lanczos analysis methods and others~\cite{Ostmeyer:2024qgu,Chakraborty:2024scw}.
    In this work, we note another coincidence: the Lanczos algorithm is equivalent to the well-known Rayleigh-Ritz method applied to Krylov subspaces.
    Rayleigh-Ritz provides optimal eigenvalue approximations within subspaces; we find that spurious-state filtering allows these optimality guarantees to be retained in the presence of statistical noise.
    We explore the relation between Lanczos and Prony's method, their block generalizations, generalized pencil of functions (GPOF), and methods based on the generalized eigenvalue problem (GEVP), and find they all fall into a larger ``Prony-Ritz equivalence class'', identified as all methods which solve a finite-dimensional spectrum exactly given sufficient correlation function (matrix) data.
    This equivalence allows simpler and more numerically stable implementations of (block) Lanczos analyses.
\end{abstract}

\maketitle

\section{Introduction}
\label{sec:intro}

Inverse problems, in which a finite number of measurements are used to constrain an unknown function from which the data were sampled, are ubiquitous in science and have been studied for centuries.
In 1795, Gaspard de Prony showed that inverse problems involving reconstruction of a time series as a sum of exponentials, including inverse Fourier transforms and inverse Laplace transforms, can be solved exactly if the number of data points is equal to the number of unknowns in the sum~\cite{Prony}.
Prony's method has a long history of successful applications to science, engineering, and signal processing~\cite{Auton:1981}.

This reconstruction problem arises in lattice quantum chromodynamics (LQCD), where correlation functions (correlators) are calculated in the time domain and used to constrain the spectrum of energies formally related by an inverse Laplace transform.
Prony's method was first applied to LQCD two decades ago~\cite{Fleming:2004hs,Lin:2007iq,Fleming:2009wb,Beane:2009kya}, where it was found to converge faster than simple ``effective masses'' formed from finite differences of logs of adjacent correlator points.
However, applications to long time series in LQCD were obstructed by statistical noise leading to unphysical results including complex energies when $\gtrsim 4$ exponentials were included~\cite{Fleming:2009wb,Beane:2009kya,Cushman:2019tcv}.
Other reconstruction techniques previously applied to signal processing such as the generalized pencil-of-functions (GPOF) method~\cite{Hua:1989,Sarkar:1995} were subsequently applied to LQCD correlators and found to lead to analogous convergence benefits and similar limitations~\cite{Aubin:2010jc,Aubin:2011zz,Green:2014xba,Schiel:2015kwa,Ottnad:2017mzd,Fischer:2020bgv}.
A block Prony algorithm extending Prony's method to matrix-valued time series was also developed and applied to LQCD~\cite{Fleming:2023zml}.

The Rayleigh-Ritz method~\cite{Ritz:1909} was introduced by Walther Ritz in 1909 in order to approximate solutions to boundary-value problems in classical mechanics.
Its application to Hamiltonians in quantum mechanics leads to the variational principle and methods to approximate energy spectra of atoms and nuclei by diagonalizing matrix representations of the Hamiltonian within finite subspaces of Hilbert space~\cite{MacDonald:1933}.
The eigenvalue approximations provided by the Rayleigh-Ritz method are globally optimal approximations to the true spectrum of the Hamiltonian given only the knowledge of a specific subspace~\cite{Parlett}.
When applied to a Krylov space iteratively constructed from an initial vector, the Rayleigh-Ritz method is equivalent to the Lanczos algorithm~\cite{Lanczos:1950zz}.

In recent months, an alternative correlator analysis method based on applying the (scalar or block) Lanczos algorithm to a time-evolution operator has been proposed~\cite{Wagman:2024rid,Hackett:2024xnx,Hackett:2024nbe} and applied to LQCD correlators~\cite{Wagman:2024rid,Hackett:2024xnx,Ostmeyer:2024qgu,Chakraborty:2024scw,Hackett:2024nbe,Zhang:2025hyo}.
This method again features the appearance of unphysical energies when applied to long time series of noisy data.
However, there is a long history of mathematical research on roundoff errors in linear algebra applications of Lanczos~\cite{Lanczos:1950zz,Paige:1971,Parlett}.
It has been understood that ``spurious eigenvalues'' arise as signals of convergence of a particular approximate eigenvalue within working precision. Methods such as selective reorthogonalization~\cite{Parlett:1979} and the Cullum-Willoughby test~\cite{Cullum:1981,Cullum:1985} have been developed to prevent or retroactively filter out spurious eigenvalues, respectively.

A generalization of the Cullum-Willoughby test to noisy Monte Carlo estimates of correlation functions was introduced in Ref.~\cite{Wagman:2024rid}.
The Cullum-Willoughby test was shown to be formally equivalent to a physically motivated ``ZCW test'' introduced in Ref.~\cite{Hackett:2024nbe}, in which states that have spuriously small initial-state overlap are identified as artifacts of noise. 
The Cullum-Willoughby and ZCW tests provide mathematically well-understood and physically motivated tests for filtering between physical energy approximations and noise artifacts in the Lanczos framework.
This provides a reconstruction method applicable to long time series that achieves faster ground-state convergence than effective masses, while at the same time avoiding the exponential degradation of signal-to-noise ratios (SNR) arising for the latter.
Filtered Lanczos results further provide excited-state energy approximations and rigorous two-sided bounds on differences between exact and approximate energies.

Ref.~\cite{Fischer:2020bgv} found that Prony's method is equivalent to GPOF, there called the ``Prony Generalized Eigenvalue Method'' (PGEVM) based on its relation to generalized eigenvalue problem (GEVP) methods.\footnote{The steps of applying PGEVM---solving a GEVP involving a pair of Hankel matrices---are the same as the steps of GPOF. 
Equivalence between PGEVM and GPOF was noted already in Ref.~\cite{Fischer:2020bgv}, although the applicability of GPOF to generic signal processing problems~\cite{Hua:1989,Sarkar:1995} is not appreciated in that work.}
Ref.~\cite{Wagman:2024rid} noted that unfiltered (scalar) Lanczos results were numerically identical to results from Prony's method.
Both Refs.~\cite{Ostmeyer:2024qgu} and \cite{Chakraborty:2024scw} showed the equivalence of (scalar) Lanczos and GPOF, 
called PGEVM in Ref.~\cite{Ostmeyer:2024qgu} and Transfer-matrix GEVP (TGEVP) in Ref.~\cite{Chakraborty:2024scw}.
Equivalences between GEVP correlator analysis methods~\cite{Luscher:1990ck,Blossier:2009kd} and block Lanczos were shown in Ref.~\cite{Hackett:2024nbe}, where further equivalences with block Prony and GPOF were noted.
These thus-far piecemeal identifications suggest some underlying structure.

In this work, we identify that underlying structure to provide a general unifying picture of correlator-analysis methods.
In particular, we find that Prony methods, GPOF, and Lanczos fall into a broader equivalence class with a simple defining criterion: it includes any method that exactly reconstructs a finite-dimensional spectrum given sufficient input data.
All methods in this equivalence class will produce identical outputs when applied to any correlator (matrix), before any postprocessing to treat noise effects.

Furthermore, these methods may all be viewed as applying the Rayleigh-Ritz (RR) method to approximate the spectrum of a ``transfer matrix'' describing the time-series evolution of the data.
When applied to lattice QCD correlators, part of this eigenspace may be identified with a subspace of the Hilbert space of physical states.
This Hilbert-space description, which is manifest from the Lanczos perspective, provides a useful theoretical starting point for applying filtering methods that identify and remove ``spurious states'' arising from noise that have unphysical properties~\cite{Wagman:2024rid,Hackett:2024xnx,Hackett:2024nbe}.
However, it is not necessary to use the Lanczos iteration in order to apply spurious-state filtering---our equivalence proof implies that any method that exactly diagonalizes some transfer-matrix approximation can be interpreted as indirectly applying RR in a subspace of Hilbert space.
Technologies developed for Lanczos applications such as residual bounds~\cite{Parlett}, Kaniel-Paige-Saad (KPS) convergence theory~\cite{Kaniel:1966,Paige:1971,Saad:1980}, and spurious-state filtering~\cite{Wagman:2024rid,Hackett:2024xnx,Hackett:2024nbe} can therefore be applied to results from Prony's method, GPOF, GEVP methods, and even effective masses.

Before proceeding, we first provide a summary of the main results of this work.
To set the stage, note that a set of $2m$ complex $r \times r$ matrices $\aset{\bm{C}(t) | \forall_{t \in \{0,\ldots,2m-1\}}}$ may admit a decomposition of the form
\begin{equation}\label{eq:block-spectral-sum}
    \bm{C}(t) = \sum_{k=0}^{rm-1} \bm{a}_k \lambda_k^t ,
\end{equation}
in terms of $d=rm$ complex eigenvalues $\lambda_k$ and rank-one amplitude matrices $\bm{a}_k$.\footnote{This means that $\bm{a}_k$ can be represented by an outer product of overlap factors $a_{k,ab} = Z^{R*}_{ka} Z^L_{kb}$. The overlap factors are individually defined only up to an overall normalization and phase ambiguity, but these cancel such that $a_{k,ab}$ is uniquely defined.}
The scalar case corresponds to the block case with $r=1$, and therefore it is sufficient to state general results for the block case.
We assume the decomposition exists and that the eigenvalues $\lambda_k$ are non-degenerate.

We define the Prony-Ritz equivalence class as any method taking $\aset{\bm{C}(t) | \forall_{t \in \{0,\ldots,2rm-1\}}}$ as inputs and providing as outputs $rm$ eigenvalues $\lambda_k'$ and amplitudes $\bm{a}_k'$ satisfying $\bm{C}(t) = \sum_k \bm{a}_k' (\lambda_k')^t$ as in \cref{eq:block-spectral-sum}.
Equivalence between the results of any such methods necessarily holds when the decomposition \cref{eq:block-spectral-sum} is unique and thus $\lambda_k'=\lambda_k$, $\bm{a}_k'=\bm{a}_k$.
We find that uniqueness holds under weak conditions that may be expected to be satisfied by any noisy correlator measurement.

To provide the uniqueness conditions precisely, we must first define the $(m,r) \times (m,r)$ (block) Hankel matrices,
\begin{equation}\label{eq:hankel}
\begin{split}
    &\bm{H}^{(p)}_{st} \equiv \bm{C}(p+s+t) \\
    &= \begin{bmatrix}
        \bm{C}(p)     & \bm{C}(p+1)  & \cdots & \bm{C}(p+m-1) \\
        \bm{C}(p+1)   & \bm{C}(p+2)  & \cdots & \bm{C}(p+m)   \\
        \vdots   & \vdots  & \ddots & \vdots   \\
        \bm{C}(p+m-1) & \bm{C}(p+m)  & \cdots & \bm{C}(p+2m-2)
    \end{bmatrix}_{st}
    .
\end{split}
\end{equation}
As shown below, the correlator decomposition is unique---and thus equivalence between any methods in the RR class must hold---whenever $\bm{H}^{(0)}$ is invertible.
In the block case, we find a second condition on the rank of a Vandermonde-like matrix must also hold; in the scalar case $r=1$, this second condition is automatically satisfied by the assumed non-degeneracy of the $\lambda_k$.
When these conditions hold, the $\lambda_k$ and $\bm{a}_k$ can be conveniently obtained by diagonalizing $[\bm{H}^{(0)}]^{-1} \bm{H}^{(1)}$ or solving the equivalent GEVP; the latter defines the GPOF algorithm~\cite{Hua:1989,Sarkar:1995}.

The remainder of this work proceeds as follows.
\Cref{sec:rr-theory} provides theoretical background including Hilbert-space definitions of the Rayleigh-Ritz method, block Krylov spaces, their oblique generalizations, and spurious state filterin.
\Cref{sec:rr-practice} presents how these methods may be implemented in analyses of correlator (matrix) data, both in general and in the special case where they may be reframed as a GEVP.
\Cref{sec:convergence} discusses the different species of convergence behaviors that characterize different methods for correlator analysis.
\Cref{sec:prony-uniqueness} proves the uniqueness theorem stated above, in the process reviewing the derivation of Prony's method and its block generalization.
\Cref{sec:prony-ritz} discusses important limitations and subtleties in this equivalence, and provides a complete taxonomy of correlator analysis methods from the RR perspective (see \Cref{tab:tax}).
Conclusions are presented in \cref{sec:conclusions}.

Note that throughout this work, all sums are denoted explicitly and repeated indices \emph{do not} imply summation.
\Cref{sec:glossary} provides further notes on notation and an extensive glossary of terms.

\section{Rayleigh-Ritz Theory}
\label{sec:rr-theory}

This section briefly reviews the Rayleigh-Ritz method for approximating eigenvalues of Hermitian matrices in a Hilbert-space framework. It then describes its oblique generalization, which is necessary to treat non-Hermitian matrices with distinct left and right eigenvectors.

\subsection{Rayleigh quotients}

For a quantum theory with Hamiltonian $H$, the ground-state energy is---by definition---the lowest energy that can be obtained by any state $\ket{\psi}$ in the full Hilbert space $\mathcal{H}$, i.e.,\footnote{Note that the quotient in \cref{eq:var} is only well-defined if $\ket{\psi} \neq 0$. All vectors in this section are assumed to be non-zero unless otherwise stated.}
\begin{equation}\label{eq:var}
  E_0 = \min_{\ket{\psi} \in \mathcal{H}} \frac{ \braket{\psi | H | \psi} }{\braket{\psi | \psi} }.
\end{equation}
For a generic Hermitian operator $H$, Eq.~\eqref{eq:var} is the statement that the smallest eigenvalue of $H$ minimizes the \emph{Rayleigh quotient} $\braket{\psi | H | \psi} / \braket{\psi | \psi}$, which for unit-normalized states is simply the matrix element of $H$.
The eigenvectors $\ket{n}$ of $H$ are in one-to-one correspondence with the stationary points of the Rayleigh quotient and the entire spectrum of eigenvalues can be obtained from the Courant-Fischer min-max theorem (see Ref.~\cite{Parlett}, Chapter 10-2),
\begin{equation}
  E_n = \min_{\mathcal{S}^{(n+1)}\subset \mathcal{H}} ~ \max_{\ket{\psi} \in \mathcal{S}^{(n+1)}} ~ \frac{ \braket{\psi | H | \psi} }{ \braket{\psi | \psi} },
\end{equation}
where $\mathcal{S}^{(n)} \subset \mathcal{H}$ is any $n$-dimensional subspace of $\mathcal{H}$.
In words, the theorem says that $E_n$ is the minimum value of the Rayleigh quotient achievable by a set of $n+1$ vectors from which the state is chosen adversarially as the linear combination maximizing the Rayleigh quotient (with other linear combinations describing the $n$ lower-energy eigenvectors).
This is equivalent to characterizing $E_n$ as the minimum energy that can be achieved in a subspace of Hilbert space that is orthogonal to $\Span(\ket{0},\ldots,\ket{n-1})$.
In this sense, the entire energy spectrum can be characterized as constrained minima of the Hamiltonian Rayleigh quotient.
 
For a lattice field theory, a Hamiltonian cannot be constructed straightforwardly but instead a transfer matrix $T$ implements discrete time translations~\cite{Luscher:1976ms}.
The continuum relation $T = e^{-H}$ can be formally inverted to define a (non-local) lattice Hamiltonian operator as $H = -\ln T$, where here and below we use units where the lattice spacing equals unity.
Positivity of the transfer matrix holds for all of Hilbert space in relatively simple lattice gauge theories~\cite{Luscher:1976ms} and more generally for the physically interesting sector of states whose energies are small compared to the lattice cutoff~\cite{Luscher:1984is}.
After restricting to this sector if needed, the eigenvalues of the $T$ are positive-definite and can be ordered as $\lambda_0 > \lambda_1 > \ldots$. 
The spectrum of lattice energies is then defined as $E_n \equiv - \ln \lambda_n$.

The transfer matrix eigenvalues satisfy an analog of the Courant-Fischer min-max theorem,
\begin{equation}\label{eq:max-min}
  \lambda_n = \max_{\mathcal{S}^{(n+1)} \in \mathcal{H}} ~ \min_{\ket{\psi} \in \mathcal{S}^{(n+1)}} ~ \frac{ \braket{\psi | T | \psi} }{ \braket{\psi | \psi} },
\end{equation}
where max and min have exchanged roles because small $\lambda_n$ corresponds to large $E_n$.
The energy spectrum of a lattice theory can therefore be characterized as constrained maxima of the transfer-matrix Rayleigh quotient.

Determining the eigenvalues of $T$ can be reframed as a GEVP on the transfer matrix in the (non-orthogonal) basis $\aset{\ket{\xi_i}}$.
To see this, first note that, generically, any GEVP $(\bm{M}, \bm{G})$ with Hermitian $\bm{M}$ and positive-definite $\bm{G}$,
\begin{equation}
  \bm{M} \vec{g}_n = \bm{G} \vec{g}_n \lambda_n,
\end{equation}
where $\lambda_n$ and $\vec{g}_n$ are the generalized eigenvalues and eigenvectors, respectively, satisfies a min-max theorem similar to that of \cref{eq:max-min}~\cite{stewart1979pertubation,householder2013theory}:
\begin{equation}\label{eq:gevp_max-min}
    \lambda_n = \max_{\mathcal{C}^{(n+1)} \subset \mathbb{C}^{\text{dim}(\mathcal{H})}} ~
    \min_{\vec{g} \in \mathcal{C}^{(n+1)}} ~
    \frac{\vec{g}^\dag \bm{M} \vec{g}}
    {\vec{g}^\dag \bm{G} \vec{g}} ,
\end{equation}
where $\mathcal{C}^{(n)}$ is an $n$-dimensional subspace of $\mathbb{C}^{\text{dim}(\mathcal{H})}$.
Now, note that we can obtain \cref{eq:gevp_max-min} directly from \cref{eq:max-min} by fixing some basis set $\{\ket{\RRvec_i}\}$ and taking $\ket{\psi} = \sum_i g_i \ket{\RRvec_i}$ for $g_i \in \mathbb{C}^{\text{dim}(\mathcal{H})}$.
Identifying
\begin{equation}
    [\bm{M}]_{ij} = \braket{\RRvec_i | T | \RRvec_j} ~~\,\text{and}~~~
    [\bm{G}]_{ij} = \braket{\RRvec_i | \RRvec_j} ~ ,
\end{equation}
we recover \cref{eq:gevp_max-min} exactly.
This demonstrates that $T$ eigenvalues and eigenvectors can be obtained via a GEVP, even in the situation that the matrix elements of $T$ are only known in some non-orthogonal basis, as will be discussed in the context of correlator analysis in Sec.~\ref{sec:gevp}, below.

\subsection{Ritz values and vectors}
\label{subsec:ritz}

We cannot in practice optimize over the infinite-dimensional Hilbert spaces arising for lattice gauge theories in order to compute the exact spectrum.
Therefore, define a finite $d$-dimensional subspace $\mathcal{S}^{(d)} \subset \mathcal{H}$,
\begin{equation}
  \mathcal{S}^{(d)} \equiv \Span(\ket{\xi_1},\ldots,\ket{\xi_d}),
\end{equation}
where the $\ket{\xi_i}$ with $i \in \{1,\ldots,d\}$ represent a set of computationally accessible states for which the matrix elements $\braket{ \xi_i | T | \xi_j }$ can be calculated.
It is convenient to introduce an equivalent orthonormal basis $\{ \ket{v_1},\ldots,\ket{v_d} \}$ satisfying $\braket{v_i | v_j} = \delta_{ij}$ as well as $\mathcal{S}^{(d)} =  \Span(\ket{v_1},\ldots,\ket{v_d})$, e.g.~through the Gram-Schmidt process.
Matrix elements of $T$ in such an orthonormal basis are denoted
\begin{equation}
    T^{(d)}_{ij} \equiv \braket{v_i | T | v_j}.
\end{equation}
The Lanczos algorithm explicitly constructs such an orthonormal basis, the Lanczos basis, in which $T^{(d)}_{ij}$ takes a particularly simple tridiagonal form; here it is not assumed that $\{ \ket{v_1},\ldots,\ket{v_d} \}$ is the Lanczos basis.

Subspace projection is achieved through the operator
\begin{equation}\label{eq:Pm}
  \Pm{d} \equiv \sum_{j=1}^d \ket{v_j} \bra{v_j} ,
\end{equation}
which is a projection operator satisfying $[\Pm{d}]^2 = \Pm{d}$ by the orthonormality of the $\ket{v_j}$.
The projection of $T$ to the computationally accessible subspace is
\begin{equation}
  T^{(d)} \equiv \Pm{d} T \Pm{d} 
  = \sum_{ij} \ket{v_i} T^{(d)}_{ij} \bra{v_j} ~ .
\end{equation}
It is convenient to define an extension of the orthonormal basis $\{\ket{v_1},\ldots,\ket{v_d}\}$ for $\mathcal{S}^{(d)}$ to a (usually infinite) basis $\{\ket{v_1},\ldots,\ket{v_d},\ket{v_{d+1}},\ldots\}$ for $\mathcal{H}$.
In this extended basis, the matrix elements of $T^{(d)}$ take the form
\begin{equation}
\begin{split}
  T^{(d)}_{ij} &\equiv \braket{ v_i | T^{(d)} | v_j } \\
  &= \begin{cases} \braket{ v_i | T | v_j }, & 1 \leq i,j \leq d \\ 0, & \text{otherwise} \end{cases},
  \end{split}
\end{equation}
The matrix $T^{(d)}_{ij}$ with $1 \leq i,j \leq d$ can be viewed as a $d \times d$ sub-matrix of the full infinite-dimensional transfer matrix $\braket{ v_i | T | v_j }$.

The \emph{Ritz values} are the eigenvalues of $T^{(d)}$, denoted $\lambda_n^{(d)}$.
By the max-min theorem, they satisfy\footnote{Note $\mathcal{S}^{(d)} = \Span(\aset{\ket{\xi_i}})$ has taken the role of $\mathcal{H}$ in \cref{eq:max-min}.}
\begin{equation}\label{eq:ritz}
  \begin{split}
    \lambda_n^{(d)} &= \max_{\mathcal{S}^{(n+1)} \subset \mathcal{S}^{(d)}} ~ \min_{\ket{\psi} \in \mathcal{S}^{(n+1)}  } \frac{ \braket{\psi | T^{(d)} | \psi} }{ \braket{\psi|\psi}} \\
   &= \max_{\mathcal{S}^{(n+1)} \subset \mathcal{S}^{(d)}} ~ \min_{\ket{\psi} \in \mathcal{S}^{(n+1)}  } \frac{\braket{\psi |\Pm{d} T \Pm{d}| \psi}}{\braket{\psi|\psi}}.
  \end{split}
\end{equation}
Therefore $\lambda_0^{(d)}$ is the maximum value of the Rayleigh quotient of $T$ that can be achieved in any state that has been projected to the subspace $\mathcal{S}^{(d)}$.
Other Ritz values are other appropriately constrained maxima.

Put another way, the max-min theorem shows that the Ritz values are the optimal approximations to $T$ eigenvalues achievable within $\mathcal{S}^{(d)}$ in the sense that they are (constrained) maxima of its projection to the subspace.
Further senses in which the Ritz values define optimal subspace eigenvalue approximations are discussed in Ref.~\cite{Parlett}, Chapter 11-4.

The \emph{Ritz vectors} are the corresponding $T^{(d)}$ eigenvectors $\ket{y_k^{(d)}}$ satisfying
\begin{equation}\begin{aligned}
  T^{(d)} \ket{y_k^{(d)}} &= \ket{y_k^{(d)}} \lambda_k^{(d)} ~ ,
  \\
  \bra{y_k^{(d)}} T^{(d)} &= \lambda_k^{(d)} \bra{y_k^{(d)}}  ~ .
\end{aligned}\end{equation}
The Ritz vectors are not exact eigenvectors of $T$; however, it is the case that $T \ket{ y_k^{(d)} }$ is equal to $\ket{y_k^{(d)}} \lambda_k^{(d)}$ plus a remainder term that is orthogonal to $\mathcal{S}^{(d)}$~\cite{Parlett}.

The Ritz values and vectors can also be identified with the GEVP solutions obtained from the subspace analog to \cref{eq:gevp_max-min}.
With $\{ \ket{\RRvec_i} \}$ denoting a (generically non-orthogonal) basis for $\mathcal{S}^{(d)}$, this takes the form
\begin{equation}\label{eq:max-minGEVP}
    \lambda_n^{(d)} = \max_{\mathcal{C}^{(n+1)} \subset \mathbb{C}^{d}} ~
    \min_{\vec{g} \in \mathcal{C}^{(n+1)}}
    \frac{\vec{g}^\dag \bm{M} \vec{g}}
    {\vec{g}^\dag \bm{G} \vec{g}} ,
\end{equation}
where $[\bm{G}]_{ij} = \braket{\RRvec_i | \RRvec_j}$ as before and 
\begin{equation}
\begin{aligned}
    [\bm{M}]_{ij} &= \braket{\RRvec_i | T^{(d)} | \RRvec_j} =  \braket{\RRvec_i | \Pm{d} T \Pm{d} | \RRvec_j} \\
    &= \braket{\RRvec_i | T | \RRvec_j},
\end{aligned}
\end{equation}
can be defined exactly as in \cref{eq:gevp_max-min} since $\Pm{d}  \ket{\RRvec_i} = \ket{\RRvec_i}$ by construction.
Upon identifying correlator data with elements of $\bm{G}$ and $\bm{M}$ below, this result is sufficient to establish that the Ritz values are identical to the generalized eigenvalues of the GEVP ($\bm{H}^{(0)}$, $\bm{H}^{(1)}$) for the Hankel matrices (\cref{eq:hankel}) when $T$ is Hermitian and positive-definite.

To discuss further properties of the Ritz values and generalize these definitions to non-Hermitian $T$ capable of describing generic complex correlator matrices, we next discuss vector spaces induced by the action of $T,T^2,\ldots,T^t$ in more detail.

\subsection{Oblique Rayleigh-Ritz}

The ORR method approximates some of the eigenvalues of an operator $T$ as the eigenvalues of the projection of that operator onto left and right subspaces spanned by two distinct sets of vectors.
Although less well-known than RR, ORR has been studied in particular in the context of oblique Lanczos algorithms~\cite{Saad:1981,Saad:1982,Saad:2011}.
The precise definition of ORR proceeds as follows.

Suppose we have two sets of vectors, $\{\ket{\RRvec^R_i}\}$ and $\{\bra{\RRvec^L_j}\}$, each of length $d$.\footnote{The generalization of ORR to sets of left- and right-vectors with different lengths is more complicated; see e.g.~the matrix-Prony method proposed in Ref.~\cite{Beane:2009kya} whose connection to ORR with rectangular $\bm{G}$ and $\bm{M}$ is discussed in \cref{subsec:tax}.}
Suppose also we are able to compute the matrix elements
\begin{equation}
    [\bm{G}]_{ij} \equiv \braket{\RRvec^L_i | \RRvec^R_j}
    ~~\,\text{and}~~~
    [\bm{M}]_{ij} \equiv \braket{\RRvec^L_i | T | \RRvec^R_j}.
\end{equation}
In full generality the $d \times d$ matrices $\bm{G}$ and $\bm{M}$ may be complex, but are typically real in applications to correlator analyses.
We may construct biorthonormal left and right bases by first decomposing $\bm{G}$ as
\begin{equation}\label{eq:orr-G-decomp}
    \bm{G} = \bm{\RRmat}^L \bm{\RRmat}^R,
\end{equation}
then defining
\begin{equation}\label{eq:orr-vR-vL-def}
\begin{aligned}
    \ket{v^R_i} &= \sum_j \ket{\RRvec^R_j} [(\bm{\RRmat}^R)^{-1}]_{ji},
    \\
    \bra{v^L_i} &= \sum_j [(\bm{\RRmat}^L)^{-1}]_{ij} \bra{\RRvec^L_j},
\end{aligned}\end{equation}
such that
\begin{equation}\begin{aligned}
    \braket{v^L_i | v^R_j} 
    &= \sum_{kl} [(\bm{\RRmat}^L)^{-1}]_{ik} \braket{\RRvec^L_k | \RRvec^R_l} [(\bm{\RRmat}^R)^{-1}]_{lj}
    \\
    &= [(\bm{\RRmat}^L)^{-1} \bm{\RRmat}^L \bm{\RRmat}^R (\bm{\RRmat}^R)^{-1} ]_{ij}
    = \delta_{ij},
\end{aligned}\end{equation}
are biorthonormal as desired.

Any choice of invertible $\bm{\RRmat}^R$ and $\bm{\RRmat}^L$ satisfying \cref{eq:orr-G-decomp} will give equivalent results for all physical quantities, including Ritz values and any other Ritz-vector matrix elements.
Different choices of how to perform this decomposition define different ``oblique conventions''.
In applications of interest, $\bm{G}$ will be non-singular but may have non-positive eigenvalues, so it is not generically possible to take $\bm{\RRmat}^R = [\bm{\RRmat}^L]^\dagger$.
An oblique construction with distinguished right and left bases is thus necessary in practice.

The biorthonormal vectors $\{\ket{v^R_j}\}$ and $\{\ket{v^L_j}\}$ can be used to define an oblique projection operator,
\begin{equation}
    \mathcal{P}^{(d)} = \sum_j \ket{ v^R_j }\bra{ v^L_j } ~ ,
\end{equation}
and the subspace approximation to the transfer matrix $T^{(d)} = \mathcal{P}^{(d)} T \mathcal{P}^{(d)}$. Its matrix elements can be computed concretely as
\begin{equation}\label{eq:rr-Tij-def}
\begin{aligned}
    T^{(d)}_{ij} &\equiv \braket{v^L_i | T^{(d)} | v^R_j} = \braket{v^L_i | T | v^R_j} 
    \\
    &= \sum_{kl} [(\bm{\RRmat}^L)^{-1}]_{ik} \braket{\RRvec^L_k | T | \RRvec^R_l} [(\bm{\RRmat}^R)^{-1}]_{li}
    \\
    &= [(\bm{\RRmat}^L)^{-1} \bm{M} (\bm{\RRmat}^R)^{-1} ]_{ij}.
\end{aligned}\end{equation}
This allows carrying out the eigendecomposition of the operator
\begin{equation}
    T^{(d)} = \sum_{k=0}^{d-1} \ket{y^{R(d)}_k} \lambda^{(d)}_k \bra{y^{L(d)}_k} ~ ,
\end{equation}
which yields Ritz values as well as the necessary information to compute Ritz-vector matrix elements.
Estimators for quantities like overlaps $\braket{\xi^{R/L}_i | y^{R/L(m)}_k}$ and matrix elements $\braket{y^{L(m)} | J | y^{R(m)}}$ of generic operators $J$ follow directly; explicit forms are deferred to \cref{sec:rr-practice} where they may be better understood in the context of correlator analyses.

\subsection{Block Krylov spaces}
\label{sec:block-krylov}

The \emph{Krylov space} $\mathcal{K}^{(m)}(\ket{\psi})$ of the transfer matrix $T$ and a vector $\ket{\psi}$ is defined as
\begin{equation}
  \mathcal{K}^{(m)}(\ket{\psi}) \equiv \Span(\ket{\psi}, T\ket{\psi}, T^2 \ket{\psi}, \ldots, T^{m-1} \ket{\psi}) .
\end{equation}
As long as $\ket{\psi}$ is not an eigenvector of $T$, then $\mathcal{K}^{(m)}(\ket{\psi})$ will be an $m$-dimensional subspace of $\mathcal{H}$.
The Krylov vectors $\ket{k_t} \equiv T^t \ket{\psi}$ with $t \in \{0,\ldots,m-1\}$, provide a non-orthogonal basis for $\mathcal{K}^{(m)}(\ket{\psi})$.
We may take the Krylov subspace as the subspace for RR, i.e., take $\aset{\ket{\xi_i}} = \aset{\ket{k_t}}$.
An equivalent orthonormal basis $\{ \ket{v_1}, \ldots, \ket{v_m} \}$ satisfying $\braket{v_i | v_j} = \delta_{ij}$ and $\mathcal{K}^{(m)}(\ket{\psi}) =  \Span(\ket{v_1},\ldots,\ket{v_m})$ can be used to define the Krylov-space projection operator by Eq.~\eqref{eq:Pm} just as in the generic subspace case.

The transfer matrix projected to Krylov space is denoted $T^{(m)} = \Pm{m} T \Pm{m}$ and its eigenvalues, the Ritz values, are denoted $\lambda_n^{(m)}$ as in the general case above.
By the max-min theorem, \cref{eq:max-min}, the Ritz values provide constrained maxima of $T$ projected to Krylov space.

Given an $r$ dimensional subspace $\mathcal{S}^{(r)}$ instead of simply an initial vector $\ket{\psi}$, it is possible to define an $rm$-dimensional block Krylov space~\cite{Gutknecht:2009,Guennouni:2002,Musco:2015}
\begin{equation}
  \mathcal{K}^{(rm)}(\mathcal{S}^{(r)}) \equiv \mathcal{K}^{(m)}(\ket{\psi_0}) \oplus \ldots \oplus \mathcal{K}^{(m)}(\ket{\psi_{r-1}}).
\end{equation}
A basis of block Krylov vectors $\ket{k_{ta}} \equiv T^t \ket{\psi_a}$, equivalent orthonormal vectors $\ket{v_{ja}}$ satisfying $\braket{v_{ia} | v_{jb}} = \delta_{ij}\delta_{ab}$ with $a,b \in \{0,\ldots,r-1\}$ and $j \in \{1,\ldots,m\}$, and projection operators 
\begin{equation}
    \Pm{rm} \equiv \sum_{j=1}^{m} \sum_{a=0}^{r-1} \ket{v_{ja}}\bra{v_{ja}},
\end{equation}
are defined analogously to the scalar case.
The Ritz values for a block Krylov space are again defined as eigenvalues of 
\begin{equation}\label{eq:Tproj}
    T^{(rm)} \equiv \Pm{rm} T \Pm{rm},
\end{equation}
The Ritz values $\lambda_k^{(rm)}$ and Ritz vectors $\ket{y_k^{(rm)}}$ with $k \in \{0,\ldots,rm-1\}$ are eigenvalues and eigenvectors of $T^{(rm)}$.
The Ritz values provide constrained maxima of Rayleigh quotients of the transfer matrix projected to the $rm$-dimensional block Krylov space.

\subsection{Oblique Block Krylov spaces}

When considering noisy Monte Carlo estimates, finite-statistics data typically do not correspond to time series generated by a Hermitian $T$.
Describing finite-statistics data as Krylov-space matrix elements is possible but requires the introduction of a non-Hermitian $T$~\cite{Wagman:2024rid,Hackett:2024xnx,Hackett:2024nbe}.
This motivates the definition of oblique Krylov (OK) and oblique block Krylov (OBK) spaces in this section.

A pair of vectors $\ket{\psi^R}$ and $\ket{\psi^L}$ can be used to define the \emph{oblique Krylov space} with 
\begin{equation}
    \begin{split}
    \mathcal{K}^{R(m)}(\ket{\psi^R}) &\equiv \Span(\ket{\psi^R}, T\ket{\psi^R}, \ldots, T^{m-1} \ket{\psi^R}), \\
    \mathcal{K}^{L(m)}(\ket{\psi^L}) &\equiv \Span(\ket{\psi^L}, T^\dagger \ket{\psi^L}, \ldots, [T^\dagger]^{m-1} \ket{\psi^L}).
    \end{split}
\end{equation}
A pair of $r$ dimensional subspaces $\mathcal{S}^{R(r)} \equiv \Span(\ket{\psi^R_1},\ldots,\ket{\psi^R_r})$ and $\mathcal{S}^{L(r)} \equiv \Span(\ket{\psi^L_1},\ldots,\ket{\psi^L_r})$ can be used to similarly
define the \emph{oblique block Krylov space} with 
\begin{equation}
    \begin{split}
    \mathcal{K}^{R(rm)}(\mathcal{S}^{R(r)}) &\equiv \mathcal{K}^{R(m)}(\ket{\psi_0^R}) \oplus \ldots \oplus \mathcal{K}^{R(m)}(\ket{\psi_{r-1}^R}), \\
    \mathcal{K}^{L(rm)}(\mathcal{S}^{L(r)}) &\equiv \mathcal{K}^{L(m)}(\ket{\psi_0^L}) \oplus \ldots \oplus \mathcal{K}^{L(m)}(\ket{\psi_{r-1}^L}).
    \end{split}
\end{equation}
To avoid redundancy, results below are phrased for OBK spaces with results for OK spaces obtained with $r=1$.

Krylov vectors 
\begin{equation}
\begin{split}
    \ket{k^R_{ta}} &\equiv T^t \ket{\psi^R_a}, \\
    \ket{k^L_{ta}} &\equiv [T^\dagger]^t \ket{\psi^L_a},
    \end{split}
\end{equation}
with $t \in \{0,\ldots,m-1\}$ provide bases for $\mathcal{K}^{R(rm)}$ and $\mathcal{K}^{L(rm)}$ respectively.
Even if $\ket{\psi^R_a} = \ket{\psi^L_a}$, non-Hermitian $T$ gives $\ket{k^R_{ta}} \neq \ket{k^L_{ta}}$ for $t > 0$, so 
\begin{equation}
    \mathcal{K}^{R(rm)} \neq \mathcal{K}^{L(rm)}
\end{equation}
in general.
If $\ket{\psi^R_a} = \ket{\psi^L_a}$ and $T$ is invertible, then there is a natural isomorphism\footnote{Physically, this corresponds to evolving backwards in time with $T$ for $t$ steps and subsequently evolving forwards in time with $T^\dagger$ for $t$ steps, which is a trivial operation iff $T = T^\dagger$. } between $\mathcal{K}^{R(rm)}$ and $\mathcal{K}^{L(rm)}$ obtained by identifying $\ket{k^L_{ta}}$ with $[T^\dagger]^t [T^{-1}]^t \ket{k^R_{ta}}$.
If $\ket{\psi^R_a} \neq \ket{\psi^L_a}$, then  $\mathcal{K}^{R(rm)}$ and $\mathcal{K}^{L(rm)}$ are distinct $rm$-dimensional vector spaces and an OBK space is simply a collection of two $rm$-dimensional block Krylov spaces.

One can further define biorthogonal bases $\{ \ket{v^R_{ja}} \}$ and $\{ \ket{v^L_{ja}} \}$ with $j \in \{1,\dots,m\}$ and $a \in \{1,\ldots,r\}$ satisfying $\braket{v^L_{ia} | v^R_{jb} } = \delta_{ij} \delta_{ab}$ using e.g.~Gram-Schmidt biorthogonalization.
Projection operators for OBK spaces can then be defined as
\begin{equation}\label{eq:OBK_P}
    \Pm{rm} \equiv \sum_{j=1}^{m} \sum_{a=0}^{r-1} \ket{v^L_{ja}}\bra{v^R_{ja}},
\end{equation}
which satisfy idempotency $[\Pm{rm}]^2 = \Pm{rm}$ via the biorthogonality  relation $\braket{v^L_{ia} | v^R_{jb} } = \delta_{ij} \delta_{ab}$.
Ritz values $\lambda_k^{(rm)}$, as well as distinct right and left Ritz vectors $\ket{y^{R(rm)}_k}$ and  $\ket{y^{L(rm)}_k}$, with $k \in \{0,\ldots,rm-1\}$ are defined from 
$T^{(rm)} = \Pm{rm} T \Pm{rm}$ through the spectral representation
\begin{equation}\label{eq:OBK_T}
    T^{(rm)} = \sum_{k=0}^{rm-1}  \bigl| y^{L(rm)}_k \bigr> \lambda^{(rm)}_k \bigl<  y^{R(rm)}_k \bigr|,
\end{equation}
where $\braket{ y^{L(rm)}_k | y^{R(rm)}_l } = \delta_{kl}$.
The correspondence between Ritz values and stationary points of the  Rayleigh quotient is broken in the oblique case, and in particular the max-min characterization does not apply to the (generally complex) Ritz values for OBK spaces.
Uniqueness of the spectral decomposition in this case is discussed in detail in \cref{sec:prony-uniqueness}.

\subsection{Lanczos = Krylov + Rayleigh-Ritz}

The (oblique, block) Lanczos algorithm is typically defined in terms of the Lanczos process, an iteration which builds biorthogonal bases of (block) Lanczos vectors $\ket{v^{R/L}_{ja}}$ spanning left and right (block) Krylov spaces.
Along the way, the process necessarily involves computing the elements of $T$ projected to (oblique, block) Krylov spaces, which is (block) tridiagonal.
However, Parlett~\cite{Parlett} suggests an alternative and broader definition of Lanczos:
\begin{equation}
    \text{Lanczos} = \text{Krylov} + \text{Rayleigh-Ritz} = \text{K} + \text{RR} ~ ,
\end{equation}
i.e., RR applied to vector sets defining Krylov spaces.
This definition can be easily seen to be equivalent to the usual recursive definition---the recursions guarantee the Lanczos vectors provide an orthonormal basis and are thus an instance of RR, while uniqueness of the Ritz values following e.g.~from the max-min characterization guarantees that any instance of RR will provide identical results.
This alternative definition generalizes straightforwardly to the oblique and/or block cases,
\begin{equation}
\begin{split}
    \text{Oblique block Lanczos} = \text{OBK} + \text{ORR} ~ .
\end{split}
\end{equation}
As employed in \cref{sec:rr-practice}, this broader definition gives a basis-independent prescription which is equivalent to the iterative definition.
The iterative definition can be computationally advantageous in numerical settings where tridiagonality reduces storage requirements and allows use of specialized algorithms.
However, in the setting of interest for this work---analysis of correlator data---Parlett's broader definition allows simpler and more convenient definitions with identical physics outputs and improved numerical stability.

\subsection{Filtering \& Hermitian subspaces}
\label{sec:filtering}

A practical issue that arises when analyzing noisy correlator data is that some Ritz values and vectors have ``spurious'' features that can only arise due to noise---for example, complex Ritz values arise in applications to real, symmetric correlator data that is known to be described by spectral representations with positive-definite $\lambda_n$ at infinite statistics.   
Similar features arise when applying Lanczos in the presence of roundoff errors, which results in spurious eigenvalues that do not show any apparent convergence to genuine eigenvalues.

The dynamics of spurious eigenvalues is relatively well-understood in the context of roundoff errors, where it has been proven that the appearance of spurious eigenvalues is in one-to-one correspondence with loss of orthogonality between would-be orthonormal basis vectors that occurs once a Ritz value has converged to a true eigenvalue within working precision~\cite{Paige:1971,Parlett}.
The ``Lanczos phenomenon'' is sometimes used to describe the fact that this process of spurious eigenvalue production does not significantly affect the convergence of non-spurious Ritz values~\cite{Cullum:1981,Cullum:1985}.

The Lanczos phenomenon motivates the application of ``filtering'' processes that label Ritz values (and vectors) as either spurious or non-spurious.
Previous works have explored filtering in the context of Prony's method and Lanczos using clustering algorithms~\cite{Cushman:2019tcv,Chakraborty:2024scw}, and by applying the Lanczos algorithm with a bootstrap version of the Cullum-Willoughby (CW) test~\cite{Wagman:2024rid} and a closely related ZCW test with a straightforward block Lanczos generalization~\cite{Hackett:2024nbe}.  
It has been observed that outlier-robust estimators applied to Ritz values obtained with OBKORR can provide energy estimators with asymptotically constant signal-to-noise ratios (SNRs), in contrast to power-iteration estimators whose SNR generically degrades exponentially with increasing iteration count~\cite{Wagman:2024rid,Hackett:2024xnx,Hackett:2024nbe,Ostmeyer:2024qgu,Chakraborty:2024scw}.
Although this filtering approach was developed in the context of the Lanczos iteration, we emphasize that filtering can be performed (as detailed in \cref{sec:orr-for-corr}) independently of whatever numerical implementation is employed to compute ORR estimators.

To describe filtering within the RR framework above, note that any decomposition of $\{0,\ldots,d-1\}$ into an arbitrary set $s$ and its compliment $\overline{s}$ can be used to define a decomposition of the subspace approximation to $T$,
\begin{equation}\label{eq:spur}
    T^{(d)} = \sum_{k \in s} \ket{y^{R(d)}_k} \lambda^{(d)}_k \bra{y^{L(d)}_k} + \sum_{k \in \overline{s}} \ket{y^{R(d)}_k} \lambda^{(d)}_k \bra{y^{L(d)}_k} ~ .
\end{equation}
This defines a decomposition of $\mathcal{S}^{R(d)}$ and $\mathcal{S}^{L(d)}$ into direct sums of subspaces $\mathfrak{s}^{R/L} = \Span\{ \ket{y^{R/L}_k}\, |\, k \in s \}$ and $\overline{\mathfrak{s}}^{R/L} = \Span\{ \ket{y^{R/L}_k}\, |\, k \in \overline{s} \}$.
Since this decomposition is performed in the eigenbasis of $T^{(d)}$, it trivially follows that states in $\mathfrak{s}^{R/L}$ do not mix with states in $\overline{\mathfrak{s}}^{R/L}$ under the time-evolution operation defined by the action of $T^{(d)}$.
The filtered version of $T^{(d)}$ is defined by 
\begin{equation}\label{eq:spurT}
    T^{(d)}_{\overline{\mathfrak{s}}} \equiv \mathcal{P}^{(d)}_{\overline{\mathfrak{s}}} T \mathcal{P}^{(d)}_{\overline{\mathfrak{s}}},
\end{equation}
where
\begin{equation}\label{eq:spurP}
    P^{(d)}_{\overline{\mathfrak{s}}} \equiv \sum_{k \in \overline{s}} \ket{y^{L(d)}_k} \bra{y^{R(d)}_k}.
\end{equation}
In the context of the CW test for explicit matrix applications of the Lanczos algorithm~\cite{Cullum:1981,Cullum:1985}, the set $s$ indexes the spurious eigenvalues; in Hilbert space language, $\mathfrak{s}^{R/L}$ are the spaces of spurious states~\cite{Hackett:2024xnx,Hackett:2024nbe}.
By construction, the non-spurious Ritz values correspond to the Ritz values obtained by applying ORR to $\overline{\mathfrak{s}}^{R/L}$.

A filtering prescription that is necessary in order to interpret Ritz vectors as physical states in Hilbert space is the Hermitian subspace filter introduced in Ref.~\cite{Hackett:2024xnx}.
Although the Hermitian subspace filter is often insufficient to remove all noise artifacts, it can be combined straightforwardly with the CW or ZCW test and ensures that filtered Ritz vectors can be consistently interpreted as physical states.
The Hermitian subspace filter labels states as spurious that would have unphysical negative- or complex-valued norms and/or complex-valued energies.
This provides a decomposition for a generic non-Hermitian operator $T^{(d)}$,
\begin{equation}
    T^{(d)} = \sum_{k=0}^{d_{\herm}-1} \ket{y^{(d)}_k} \lambda^{(d)}_k \bra{y^{(d)}_k} + \sum_{k=d_{\herm}}^{d-1} \ket{y^{R(d)}_k} \lambda^{(d)}_k \bra{y^{L(d)}_k} ~ ,
\end{equation}
where redundant $R/L$ labels are omitted from the first sum and the \emph{Hermitian subspace} $\mathfrak{h} \subset \mathcal{S}^{R(d)} \cap \mathcal{S}^{L(d)} \subset \mathcal{H}$ is defined to be the largest subspace spanned by Ritz vectors for which $\ket{y_l^{R(d)}} = \ket{y_l^{L(d)}}$ and $\lambda_l^{(d)} \in \mathbb{R}$, that is
\begin{equation}
\begin{split}
    \herm \equiv & ~\text{span} \left\lbrace \ket{ y^{R(d)}_l }  \forall_l \  |  \ \lambda_l^{(d)} \in \mathbb{R} \, \&  \, \ket{y^{R(d)}_l} = \ket{y^{L(d)}_l} \right\rbrace.
    \end{split}
\end{equation}
For an OBK space, it follows from this definition that
\begin{equation}
    \herm \subset \mathcal{K}^{R(rm)} \cap \mathcal{K}^{L(rm)}.
\end{equation}
The projection of $T^{(d)}$ to the Hermitian subspace, $T^{(d)}_{\herm}$, is defined by \cref{eq:spurT} with $\overline{s}$ equal to the set of indices $\aset{k}$ for which $\ket{y_k^{R(d)}} \in \herm$.
By construction $[T^{(d)}_{\herm}]^\dagger =  T^{(d)}_{\herm}$.

A Hermitian operator has $\text{dim}(\herm) = d$.
Non-Hermitian operators can have $\text{dim}(\herm) = 0$.
Intermediate cases in which $0 < \text{dim}(\herm) < d$ arise in applications of {Lanczo~(=KRR)} to matrices in the presence of roundoff~\cite{Paige:1971,Parlett:1979,Parlett,Cullum:1981,Cullum:1974} as well as LQCD correlator data~\cite{Hackett:2024xnx,Hackett:2024nbe}.
For diagonal correlator (matrix) data where\footnote{The physical interpretation of Hermitian-subspace construction and other filtering for applications with $\ket{\psi^R_a} \neq \ket{\psi^L_a}$ is more complicated and deferred to future work. } $\ket{\psi^R_a} = \ket{\psi^L_a}$, the Hermitian subspace is defined to include only real Ritz values $\lambda_k^{(d)} \in \mathbb{R}$ with degenerate right and left Ritz vectors, $\ket{y^{R(d)}_k} = \ket{y^{L(d)}_k}$ as above.

For states in the Hermitian subspace, filtered ORR inherits the optimality properties of RR for Hermitian operators.
In the context of LQCD correlator analysis, it is therefore useful to consider definitions of $s$ for which $\overline{\mathfrak{s}}^R = \overline{\mathfrak{s}}^L \subset \mathfrak{h}$ by construction.
In particular, Hermiticity of $T^{(d)}_{\herm}$ implies that filtered Ritz values with $\overline{\mathfrak{s}}^{R} = \overline{\mathfrak{s}}^{L} \subset \herm$ are equivalent to the Ritz values that would be obtained from (non-oblique) RR in the subspace spanned by  $\{ \ket{y_k^{R(d)}}\, |\, k \in \overline{s} \}$.
Such filtered OBKORR results can therefore be described simply as filtered Rayleigh-Ritz (FRR).
This equivalence means that FRR results with $\overline{\mathfrak{s}}^{R} = \overline{\mathfrak{s}}^{L} \subset \herm$ satisfy the max-min characterization of $T^{(d)}_{\overline{\mathfrak{s}}}$ eigenvalues: they are (constrained) maxima of $T$ Rayleigh quotients within the non-spurious subspace.

For asymmetric correlators, there is no Hermitian subspace even in the infinite-statistics limit because $\ket{\psi_a^R} \neq \ket{\psi_a^L}$ implies $\ket{y_k^R} \neq \ket{y_k^L}$.
Therefore, FRR results for asymmetric correlators do not inherit optimality properties of Ritz values for Hermitian operators such as the max-min characterization for subspace RR.

\section{Rayleigh-Ritz Practice}
\label{sec:rr-practice}

In this section we turn to the practical definition of the oblique Rayleigh-Ritz (ORR) method and how it may be applied to analyze correlator data.
After providing the generic definitions, we reinterpret the calculations in terms of GEVPs to provide an alternative definition without reference to intermediate bases.

\subsection{Relation to lattice QFT correlator analyses}

Lattice field theory calculations produce length $N_t$ time series of $r \times r$ correlator matrices of the form
\begin{equation}
    C_{ab}(t) = \braket{\psi^L_a(t) \psi^{R\dagger}_b(0)}
    = \braket{\psi^R_a | T^t | \psi^L_b },
\end{equation}
where $\psi^{R/L}_a$ with $a \in \{0,\ldots,r-1\}$ are interpolating operators with definite quantum numbers.
In the second equality, we have switched from Heisenberg to Schr{\"o}dinger picture via $O(t) = T^{-t} O T^t$ and assumed that thermal effects are either negligible or incorporated into the definition of $T$ as discussed in Ref.~\cite{Wagman:2024rid}.
The interpolating operators $\psi^{R/L}_a$ excite the ``initial states''
\begin{equation}
    \ket{\psi^{R/L}_a} = (\psi^{R/L}_a)^\dagger \ket{\Omega},
\end{equation}
from the vacuum $\ket{\Omega}$.
Often in practical applications $\psi^R_a = \psi^L_a$, but it is necessary to work with the general case because $R/L$ spaces must be distinguished after application of $T$ for noisy data in any case.
Convergence properties depend strongly on whether or not $\psi^R_a = \psi^L_a$ as discussed below, but application of ORR is identical.

The correlator matrix data may be reinterpreted as matrix elements involving the Krylov vectors,
\begin{equation}\label{eq:krylov_corr}
\begin{split}
    \ket{k^{R}_{ta}} &\equiv T^t \ket{\psi^{R}_a}, \\
    \ket{k^{L}_{ta}} &\equiv [T^\dagger]^t \ket{\psi^{L}_a},
    \end{split}
\end{equation}
which provide (non-orthogonal, unnormalized) basis vectors for the OBK space with $\mathcal{K}^{R(m)}$ and $\mathcal{K}^{L(m)}$ spanned by $\{ \ket{k^{R}_{ta}} \}$ and  $\{ \ket{k^{R}_{ta}} \}$, respectively, with $t \in \{0,\ldots,m-1\}$ and $a \in \{0,\ldots,r-1\}$.
The appearance of $[T^\dagger]$ in \cref{eq:krylov_corr} leads to simple relations between correlator data and matrix elements of the form
\begin{equation}
\begin{aligned}
    \braket{k^{L}_{sa} | T^p | k^R_{tb}} 
    &= \braket{\psi^L_a | T^s T^p T^t | \psi^R_b } 
    \\&= C_{ab}(s+p+t)
    = [\bm{H}^{(p)}]_{sa,tb} ~ ,
\end{aligned}
\end{equation}
for $s,t \in \{0,\ldots,m-1\}$ with $m \leq (N_t - p + 1)/2$, and where in the last equality we recognize the block Hankel matrix of \cref{eq:hankel}.

\subsection{ORR for correlator analyses}
\label{sec:orr-for-corr}

With ORR defined, the relation between correlator data and Krylov spaces established, and Parlett's broader definition of Lanczos adopted, we now proceed to derive the main results of this section.
The block Lanczos algorithm at iteration $m$ is ORR applied with the $R/L$ Krylov vectors as the $R/L$ vector sets, i.e.,
\begin{equation}
\begin{aligned}
    \aset{\ket{\RRvec^R_i} ~|~ \forall_{i \in \{1,\ldots,rm\}}} 
        &= \aset{\ket{k^R_{ta}} ~|~ \forall_{a \& t \in \{0,\ldots,m-1\}}} ~ ,
    \\
    \aset{\ket{\RRvec^L_i} ~|~ \forall_{i \in \{1,\ldots,rm\}}} 
        &= \aset{\bra{k^L_{ta}} ~|~ \forall_{a \& t \in \{0,\ldots,m-1\}}}  ~.
\end{aligned}
\end{equation}
The identification of the indices $ta$ with the single index $i$ reflects that, in contrast to the formalism of Ref.~\cite{Hackett:2024nbe}, the greater freedom of oblique conventions in ORR does not require preserving the block structure.
Given these vectors, ORR prescribes first constructing the block Hankel matrices
\begin{equation}
\begin{aligned}
    G_{sa,tb} &= \braket{ k^L_{sa} | k^R_{tb} } = C_{ab}(s+t) = H^{(0)}_{sa,tb} ~ , \\
    M_{sa,tb} &= \braket{ k^L_{sa} | T | k^R_{tb} } = C_{ab}(s+t+1) = H^{(1)}_{sa,tb} ~ ,
\end{aligned}
\end{equation}
where $(sa)$ and $(tb)$ are considered as composite indices.
Next, we decompose the $G$ matrix as\footnote{Note that each of $E^{R/L}$ are square matrices, considering $(sa)$ and $(tb)$ ``unravelled'' as composite indices.}
\begin{equation}
    G_{sa,tb} = \sum_{i=1}^{rm} E^L_{sa, i} E^R_{i,tb},
\end{equation}
using whatever choice of oblique convention, for example $\bm{\RRmat}^L = \bm{G}$ and $\bm{\RRmat}^R = \bm{1}$. 
Next applying \cref{eq:orr-vR-vL-def}, we obtain a generalization of the Lanczos vectors\footnote{In block Lanczos these vectors have additional block structure, but it is not necessary for the intermediate ORR basis to respect the block structure for ORR to obtain the same Ritz values as block Lanczos.} as 
\begin{equation}\label{eq:korr-vR-vL-def} 
\begin{aligned}
    \ket{v^R_{i}} 
    &= \sum_{tb} \ket{k^R_{tb}} [(E^R)^{-1}]_{tb,i},
    \\
    \bra{v^L_{i}} 
    &= \sum_{tb} [(E^L)^{-1}]_{i,tb} \bra{k^L_{tb}}
    .
\end{aligned}\end{equation}
Comparing with the formalism of Ref.~\cite{Hackett:2024nbe}, we may identify the inverses of $\bm{E}^{R/L}$ as a generalization of the \emph{Krylov coefficients},\footnote{
Note that this definition of the Krylov coefficients involves the interpolator basis $\{T^t \ket{\psi_a}\}$ rather than the Lanczos basis $\{T^t \ket{v_1}\}$ and are therefore equivalent to the Krylov coefficients defined in Ref.~\cite{Hackett:2024nbe} times additional factors involving the $\beta_1^{-1}$ and $\gamma_1^{-1}$ defined in that work.
}
\begin{equation}\begin{aligned}
    K^R_{ta,i} &\equiv [(E^R)^{-1}]_{ta,i} 
    \\
    K^L_{i,ta} &\equiv [(E^L)^{-1}]_{i,ta} ~ .
\end{aligned}\end{equation}

\subsubsection{Ritz values and vectors}

Computing estimators for physical quantities requires first computing the Ritz values and Ritz coefficients.
The projection of $T$ onto the R/L OBK spaces can be obtained per \cref{eq:rr-Tij-def} as\footnote{Note that Ref.~\cite{Hackett:2024nbe} uses superscripts to refer to iteration number rather than subspace dimension, i.e. $T^{(rm)}$ here is equivalent to the $T^{(m)}$ constructed from a rank-$r$ correlator matrix in Ref.~\cite{Hackett:2024nbe}. }
\begin{equation}\label{eq:Hankel_Tm}
\begin{aligned}
    T^{(rm)}_{ij}
    &= \braket{v^L_{i} | T | v^R_{j}} = [(E^L)^{-1} M (E^R)^{-1}]_{ij}
    \\
    &= \sum_{tc,sd} K^L_{i,tc} 
    \braket{k^L_{tc} | T | k^R_{sd}}
    K^R_{sd,j} 
    \\
    &= \sum_{tc,sd} K^L_{i,tc} 
    C_{cd}(t+s+1)
    K^R_{sd,j} 
    \\
    &= [\bm{K}^L \bm{H}^{(1)} \bm{K}^R]_{ij} ~ .
\end{aligned}
\end{equation}
Its eigendecomposition
\begin{equation}\label{eq:Tij-eigendecomp}
    T^{(rm)}_{ij} = \sum_k \omega^{(rm)}_{ik} \, \lambda^{(rm)}_k \, (\omega^{-1})^{(rm)}_{kj},
\end{equation}
provides the Ritz values $\lambda^{(rm)}_k$ and the change-of-basis matrices $\omega^{(rm)}_{ik}$ and $(\omega^{-1})^{(rm)}_{kj}$ required to construct right- and left-Ritz vectors as
\begin{equation}\label{eq:ritz-vec-def}
\begin{aligned}
    \ket{y^{R(rm)}_k} &\equiv \mathcal{N}_k^{(rm)} \sum_i \ket{v^R_i} \omega^{(rm)}_{ik},
    \\
    \bra{y^{L(rm)}_k} &\equiv \frac{1}{\mathcal{N}_k^{(rm)}} \sum_j (\omega^{-1})^{(rm)}_{kj} \bra{v^L_j} ~ .
\end{aligned}    
\end{equation}
Here, the normalization factors $\mathcal{N}_k$ reflect the freedom to redefine right eigenvectors by an arbitrary complex constant and that left eigenvectors, as they appear in the eigendecomposition, are uniquely defined from the right ones.
Inserting \cref{eq:korr-vR-vL-def} into \cref{eq:ritz-vec-def}, we can define the Ritz coefficients,\footnote{
Note that, as with the Krylov coefficients, these definitions of the Ritz coefficients are related to those of Ref.~\cite{Hackett:2024nbe} by additional factors involving the $\beta_1^{-1}$ and $\gamma_1^{-1}$ defined in that work.
}
\begin{equation}\begin{aligned}\label{eq:ritz_coeff_def}
    P^{R(rm)}_{ta,k} 
    &\equiv \sum_j K^R_{ta,j} \omega^{(rm)}_{jk} \mathcal{N}^{(rm)}_k,
    \\
    P^{L(rm)}_{k,ta} 
    &\equiv \sum_j \frac{1}{\mathcal{N}^{(rm)}_k} (\omega^{-1})^{(rm)}_{kj} K^L_{j,ta} ~ ,
\end{aligned}\end{equation}
which relate the Ritz vectors directly to the Krylov vectors $\aset{\ket{\RRvec^{R/L}_i}} = \aset{\ket{k^{R/L}_{ta}}}$ by
\begin{equation}
\begin{aligned}
    \ket{y^{R(rm)}_k}
    &= \sum_{ta} \ket{k^R_{ta}} P^{R(rm)}_{ta,k}
    \\
    \bra{y^{L(rm)}_k}
    &= \sum_{ta} P^{L(rm)}_{k,ta} \bra{k^L_{ta}} ~ .
\end{aligned}    
\end{equation}
As noted in \cref{sec:gevp}, Ritz coefficients may be defined and computed without explicit reference to the intermediate bases $\aset{\ket{v^{R/L}_j}}$.

We emphasize that, although this construction is substantially simpler than that of the iteration-based definitions of Ref.~\cite{Hackett:2024nbe}, it is equivalent up to a change of basis and any physical quantity computed will be identical.
This equivalence allows for values of convention-dependent quantities that could not be obtained using the iteration-based definitions: for example, $T^{(rm)}_{ij}$ is not block tridiagonal for general choices of $E^{R/L}$.
However, while the precise values of $T^{(rm)}_{ij}$ depends on the choice of oblique convention, the projection of $T$ to the R/L OBK spaces, $T^{(rm)}$, is a uniquely defined operator.
This implies that its eigenvalues, eigenvectors, and any quantities derived from them will be identical to those computed by the iteration-based formalism.

\subsubsection{Overlap factors \& normalization}

Lanczos constructions for overlap factors are straightforwardly generalized to ORR as
\begin{equation}
\begin{aligned}
    [Z^{R(rm)}_{ka}]^*
    &\equiv \braket{\psi^L_a | y^{R(rm)}_k}
    = \braket{k^L_{0a} | y^{R(rm)}_k}
    \\
    &= \sum_j \RRmat^L_{0a,j} \omega^{(rm)}_{jk} \mathcal{N}^{(rm)}_k 
    ~ ,
    \\
    Z^{L(rm)}_{ka}
    &\equiv \braket{y^{L(rm)}_k | \psi^R_a}
    = \braket{y^{L(rm)}_k | k^R_{0a}}
    \\
    &= \sum_j \frac{1}{\mathcal{N}^{(rm)}_k} (\omega^{-1})^{(rm)}_{kj} \RRmat^R_{j,0a}
    ~ .
\end{aligned}
\end{equation}
Overlaps may also be computed in terms of Ritz coefficients $P^{R/L(m)}$ as defined in \cref{eq:ritz_coeff_def} as
\begin{equation}
\label{eq:Z-factors-ritz-coeffs}
\begin{aligned}
    [Z^{R(rm)}_{ka}]^* &= \braket{\psi^L_a | y_k^{R(rm)}} = 
    \sum_{t,b} \braket{\psi^L_a | T^t | \psi^R_b} P^{R(rm)}_{tb,k} \\
    &= 
    \sum_{t,b} C_{ab}(t) P^{R(rm)}_{tb,k} 
    \\
    Z^{L(rm)}_{ka} &= \braket{y_k^{L(rm)} | \psi^R_a} = 
    \sum_{t,b} P^{L(rm)}_{k,tb} \braket{\psi^L_b | T^t | \psi^R_a}  \\
    &= 
    \sum_{t,b} P^{L(rm)}_{k,tb} C_{ba}(t) ~ .
\end{aligned}
\end{equation}

As found in Ref.~\cite{Hackett:2024nbe}, the overlap factors provide all necessary information for spurious state filtering.
Hermitian subspace filtering may be implemented for diagonal correlators with $\ket{\psi^L_a} = \ket{\psi^R_a}$ by discarding all non-real eigenvalues and applying the ``norm trick'', i.e., attempting to unit-normalize the Ritz vectors by computing
\begin{equation}\label{eq:norm_trick}
    |\mathcal{N}^{(rm)}_k|^2 \stackrel{?}{=} \frac{ \tilde{Z}^{L(rm)}_{ka} }{ \tilde{Z}^{R(rm)}_{ka} } ~,
\end{equation}
where $\tilde{Z}^{R/L}$ are unnormalized overlaps computed with $\mathcal{N}_k=1$.
If $\ket{y^{R(rm)}_k}$ and $\ket{y^{L(rm)}_k}$ are not simultaneously unit-normalizable, the value computed can be non-positive.
Such non-positivity is a contradiction, which signals that $\ket{y^{R(rm)}_k} \neq \ket{y^{L(rm)}_k}$ and that that $k$ should be discarded.
When the value computed is positive, it may be taken as $|\mathcal{N}^{(rm)}_k|^2$ to unit-normalize the $k$th Ritz vector.

Separately, the overlaps also allow filtering with the ZCW test~\cite{Hackett:2024nbe},
\begin{equation}
    \left| \sum_{ab} Z^{R(m)*}_{ka} [\bm{C}(0)^{-1}]_{ab} Z^{L(m)}_{k b} \right| < \varepsilon_{\rm ZCW} ~ ,
\end{equation}
where the cut $\epsilon_\mathrm{ZCW}$ can be chosen as discussed in Ref.~\cite{Hackett:2024nbe}.
The combination of Hermitian subspace and ZCW filtering is the complete prescription presented in Ref.~\cite{Hackett:2024nbe}.
Note that this allows spurious-state filtering to be performed without explicit reference to the intermediate bases $\aset{\ket{v^{R/L}_i}}$, as discussed further in \cref{sec:gevp}.

\subsubsection{Matrix elements}

Following the constructions presented in Refs.~\cite{Hackett:2024xnx,Hackett:2024nbe}, Ritz coefficients also allow computation of matrix elements of generic operators $J$ between Ritz vectors.
Treating $J$ which carries momentum or otherwise does not have vacuum quantum numbers requires working with distinct initial-state (unprimed) and final-state (primed) quantities.
First, ORR is applied separately to the initial-state correlator matrix $\bm{C}(t)$ to compute Ritz coefficients $P^{R(rm)}$, then to the final-state $\bm{C}'(t)$ to compute $P^{L(rm)\prime}$.
Ritz-vector matrix elements may then be computed directly from the three-point function,
\begin{equation}
\begin{aligned}
    C^\mathrm{3pt}_{ab} 
        &= \braket{\psi^{L\prime}_a(\sigma+\tau) J(\tau) \psi^{R\dagger}_b(0)}
        \\&= \braket{\psi^{L\prime}_a | T^\sigma J T^\tau | \psi^R_b}
        = \braket{k^{L\prime}_{\sigma a} | J | k^R_{\tau b}} ~ ,
\end{aligned}
\end{equation}
as
\begin{equation}
\label{eq:matrix-elements-PrPl}
\begin{aligned}
    \braket{y^{L(rm)\prime}_k | J | y^{R(rm)}_l}
    &= \sum_{\sigma a \tau b}
    P^{L(rm)\prime}_{k,\sigma a}
    \braket{k^{L\prime}_{\sigma a} | J | k^R_{\tau b}}
    P^{R(rm)}_{\tau b, l}
    \\
    &= \sum_{\sigma a \tau b}
    P^{L(rm)\prime}_{k,\sigma a}
    C^\mathrm{3pt}_{ab}(\sigma, \tau)
    P^{R(rm)}_{\tau b, l} ~ 
\end{aligned}
\end{equation}
amounting to simple matrix multiplication onto $\bm{C}^\mathrm{3pt}$.

\subsubsection{Residual bounds}

The residual bounds presented in Refs.~\cite{Wagman:2016bam,Hackett:2024xnx,Hackett:2024nbe},
\begin{equation}
    \mathrm{min}_{\lambda \in \aset{\lambda_n}} |\lambda - \lambda^{(rm)}_k|^2
    \leq |B^{R/L(rm)}_k| ~ ,
\end{equation}
where
\begin{equation}\label{eq:B-is-residual-norms}
\begin{aligned}
    B^{R(rm)}_k &\equiv \frac{
        || ~ [T - T^{(rm)}] \ket{y^{R(rm)}_k} ||^2
    }{
        \braket{y^{R(rm)}_k | y^{R(rm)}_k}
    }
    \\
    B^{L(rm)}_k &\equiv \frac{
        || \bra{y^{L(rm)}_k} [T - T^{(rm)}] ~ ||^2
    }{
        \braket{y^{L(rm)}_k | y^{L(rm)}_k}
    }
\end{aligned}
\end{equation}
offer two-sided bounds on the possible deviation of a Ritz value from a true eigenvalue.
In those works, they are defined explicitly in terms of the Lanczos iteration, with the bound at step $m$ related to Lanczos-specific quantities computed at step $m+1$.
However, their computation generalizes naturally, allowing the same (oblique convention-independent) quantities to be computed in the broader class of oblique conventions allowed by ORR over Lanczos.
In ORR, the equivalent of an additional step of Lanczos iteration is simply expanding the OBK space by addition of the next subset of vectors, $\aset{\ket{k^{R/L}_{(m+1)a}}}$.
Given this formal grounding, it is straightforward to define versions of $B^{R/L(rm)}_k$ that can be computed without reference to a Lanczos iteration.
In full generality, for an off-diagonal correlator matrix $C_{ab}(t) = \braket{\psi^L_a|T^t|\psi^R_b}$, their computation requires the corresponding diagonal correlator matrices
\begin{equation}
    \label{eq:corr-LL-RR-def}
    C^{RR}_{ab}(t) \equiv \braket{\psi^R_a| T^t |\psi^R_b}
    ~\text{ and }~
    C^{LL}_{ab}(t) \equiv \braket{\psi^L_a| T^t |\psi^L_b}
\end{equation}
from which one can define block Hankel matrices
\begin{equation}
    [\bm{H}^{(p)}_{RR}]_{st} \equiv \bm{C}^{RR}(s+t+p)
    ~\text{ and }~
    [\bm{H}^{(p)}_{LL}]_{st} \equiv \bm{C}^{LL}(s+t+p) ~ .
\end{equation}
We relegate the derivation to \cref{app:res-bounds}.
Its results are that the numerators of \cref{eq:B-is-residual-norms} may be computed as
\begin{equation}\label{eq:B-numerator}
\begin{aligned}
    &|| ~ [T - T^{(rm)}] \ket{y^{R(rm)}_k} ||^2 
    \\ & \quad
    \equiv  
        \left[ \bm{P}^{R\dagger} \bm{H}_{RR}^{(2)} \bm{P}^R \right]_{kk}
        - 2 \mathrm{Re}[\lambda^{(m)}_k] \left[ \bm{P}^{R\dagger} \bm{H}_{RR}^{(1)} \bm{P}^R \right]_{kk}
    \\ & \qquad
        + |\lambda^{(m)}_k|^2 \left[ \bm{P}^{R\dagger} \bm{H}_{RR}^{(0)} \bm{P}^R \right]_{kk}
    ~ ,
    \\
    & || \bra{y^{L(rm)}_k}  [T - T^{(rm)}] ~ ||^2
    \\ & \quad \equiv
        \left[ \bm{P}^{L} \bm{H}_{LL}^{(2)} \bm{P}^{L\dagger} \right]_{kk}
        - 2 \mathrm{Re}[\lambda^{(m)}_k] \left[ \bm{P}^{L} \bm{H}_{LL}^{(1)} \bm{P}^{L\dagger} \right]_{kk}
    \\ & \qquad
        + |\lambda^{(m)}_k|^2 \left[ \bm{P}^{L} \bm{H}_{LL}^{(0)} \bm{P}^{L\dagger} \right]_{kk} 
        ~ ,
\end{aligned}
\end{equation}
while the denominators may be computed as
\begin{equation}\label{eq:B-denominator}
\begin{aligned}
    \braket{y^{R(rm)}_k | y^{R(rm)}_k}
    &\equiv \left[ \bm{P}^{R\dagger} \bm{H}_{RR}^{(0)} \bm{P}^R \right]_{kk} ~ ,
    \\
    \braket{y^{L(rm)}_k | y^{L(rm)}_k}
    &\equiv \left[ \bm{P}^{L} \bm{H}_{LL}^{(0)} \bm{P}^{L\dagger} \right]_{kk} ~ .
\end{aligned}
\end{equation}

\subsection{Relation to GEVP: Lanczos without Lanczos vectors}
\label{sec:gevp}

Given the independence of all physical quantities from the oblique convention, a natural question to ask is whether it is possible to obtain the relevant physical quantities (energies, matrix elements, residual bounds, etc.) directly, without reference to any intermediate basis. Indeed, this can be achieved by reinterpreting the previous calculations entirely in terms of matrix elements of Krylov vectors, which exposes interesting and useful connections with GEVP and GPOF methods.

Particular choices of oblique convention allow the computation be phrased in terms of GEVPs.
For example, with the choice $\bm{E}^L = \bm{G}$ and $\bm{E}^R = \bm{1}$, solving the standard eigenproblem for $T^{(rm)} = G^{-1} M$ is equivalent to solving the (right) GEVP
\begin{equation}\label{eq:RR_GEVP}
\sum_{t b} M_{sa,tb} \, \omega_{tb,k}^{(rm)} = \lambda_k^{(rm)} \sum_{tb} G_{sa,tb} \, \omega_{tb,k}^{(rm)} ~ .
\end{equation}
Standard GEVP solvers can be applied with $(s,a)$ and $(t,b)$ flattened into indices ranging over $0,\ldots,rm-1$ so that $\bm{G}$ and $\bm{M}$ can be expressed in standard matrix form.
The generalized eigenvalues $\lambda_k^{(d)}$ are identical to the Ritz values discussed in \cref{subsec:ritz}; phrasing the problem as a GEVP thus allows their computation to be defined entirely without reference to an intermediate basis.
Determination of the Ritz values using \cref{eq:RR_GEVP} corresponds to the GPOF algorithm~\cite{Hua:1989,Sarkar:1995}: solving a GEVP of Hankel matrices $\bm{M} = \bm{H}^{(1)}$ and $\bm{G} = \bm{H}^{(0)}$.

As already noted in \cref{sec:rr-theory}, the left/right Ritz coefficients can be realized directly as the left/right generalized eigenvectors of the pair $(\bm{M}, \bm{G})$.
For the same oblique convention as \cref{eq:RR_GEVP}, the generalized eigenvectors $\omega_{jk}^{(d)}$ directly provide the right Ritz coefficients $P^{R(d)}_{ta,k}$.
For the right Ritz coefficients this can also be seen more directly via the following computation, 
\begin{equation}
\begin{aligned}
\bm{M} \bm{P}^{R}
  &=  \bm{M} \bm{K}^R \bm{\omega} \bm{\mathcal{N}} \\
  &= (\bm{K}^L)^{-1} \underbrace{\bm{K}^L \bm{M} \bm{K}^R}_{\bm{T}} \bm{\omega} \bm{\mathcal{N}} \\
&= (\bm{K}^L)^{-1} \bm{\omega} \bm{\Lambda} \bm{\mathcal{N}} \\
&= (\bm{K}^L)^{-1} (\bm{K}^R)^{-1} \bm{P}^{R} \bm{\Lambda} \\
&= \bm{E}^L \bm{E}^R \bm{P}^{R} \bm{\Lambda} \\
&= \bm{G} \bm{P}^{R} \bm{\Lambda}
\end{aligned}
\end{equation}
where
$\bm{\Lambda} \equiv \mathrm{diag}[\lambda_0, \lambda_1, \ldots, \lambda_{rm-1}]$ and  similarly $\bm{\mathcal{N}}~\equiv~\mathrm{diag}[\mathcal{N}_0, \mathcal{N}_1, \ldots, \mathcal{N}_{rm-1}]$. Here and below in this section, $(rm)$ superscripts are suppressed for clarity.
Similarly for the left Ritz coefficients:\footnote{Note that this a slightly unconventional eigenvalue equation; a left-eigenvector $\vec{g}_L$ for a pair $(\bm{M}, \bm{G})$ is typically defined to satisfy~\cite{Golub:2013}
\begin{equation}
\vec{g}_L^{\dag} \bm{M} = \lambda \vec{g}_L^{\dag} \bm{G},
\end{equation}
where $\lambda$ is the generalized eigenvalue associated to $\vec{g}_L$. \cref{eq:left-ritz-eq-eigenvectors} therefore implies that the left eigenvectors of $(\bm{M}, \bm{G})$ are the columns of $\bm{P}^{L\dag}$, rather than more typical choice of the columns of $\bm{P}^L$.}
\begin{equation}\label{eq:left-ritz-eq-eigenvectors}
\begin{aligned}
\bm{P}^{L} \bm{M} &=
  \bm{\mathcal{N}}^{-1} \bm{\omega}^{-1} \bm{K}^{L} \bm{M} \\
&= \bm{\mathcal{N}}^{-1}  \bm{\omega}^{-1} \bm{K}^{L} \bm{M} \bm{K}^{R} (\bm{K}^{R})^{-1} \\
&=\bm{\mathcal{N}}^{-1}   \bm{\omega}^{-1} \bm{T} (\bm{K}^{R})^{-1} \\
&= \bm{\mathcal{N}}^{-1}  \bm{\Lambda} \bm{\omega}^{-1} (\bm{K}^{R})^{-1} \\
&= \bm{\mathcal{N}}^{-1} \bm{\Lambda} \bm{\omega}^{-1} \bm{K}^{L} (\bm{K}^{L})^{-1} (\bm{K}^{R})^{-1} \\
&= \bm{\Lambda} \bm{P}^{L} \bm{E}^{L} \bm{E}^{R} \\
&= \bm{\Lambda} \bm{P}^{L} \bm{G}.
\end{aligned}
\end{equation}
Computing the generalized eigenvalues and eigenvectors of a matrix pencil $(\bm{M}, \bm{G})$ is a standard linear algebra routine, available in many linear algebra packages (e.g. the routine \texttt{scipy.linalg.eig} in SciPy~\cite{2020SciPy-NMeth}).

When computing Ritz coefficients as generalized eigenvectors, some additional care is needed to fix the normalization of the Ritz vectors. 
In particular, biorthonormality of the Ritz vectors must be enforced by hand.
This requires rescaling of $P^{R(rm)}_{ta,k}$ and/or $P^{L(rm)}_{k,ta}$ for each $k$ such that
\begin{equation}
    \label{eq:ritz-norm-LR}
    \braket{y^{L(rm)}_k | y^{R(rm)}_l} 
    = [\bm{P}^{L(rm)} \bm{G} \bm{P}^{R(rm)}]_{kl} 
    = \delta_{kl} ~ ,
\end{equation}
which can be achieved as long as the Ritz values are nondegenerate, as shown in \cref{app:gevp}. 
Equivalently, a consistently normalized $P^L$ may be obtained as the matrix inverse of $G P^R$.
Once \cref{eq:ritz-norm-LR} has been enforced, there still remains an additional freedom to rescale the rows of $P^{L(rm)}$ and the columns of $P^{R(rm)}$ by inverse factors, which corresponds to changing the (unphysical) normalization factor $\mathcal{N}^{(rm)}_k$.

In order to fix the normalization $\mathcal{N}^{(rm)}_k$, another useful fact proven in \cref{app:gevp} is that any right eigenvector $\vec{g}_R$ of the pair $(\bm{M}, \bm{G})$ with real eigenvalue $\lambda \in \mathbb{R}$, is also a left eigenvector with the same eigenvalue $\lambda$.
This implies that the associated left and right eigenvectors span the same subspace, and hence $P^{L(rm)*}_{k,ta} \propto P^{R(rm)}_{ta,k}$ for all $t$ and $a$ whenever $\lambda_k^{(rm)}$ is real. 
In this case, the normalization constant $\mathcal{N}^{(rm)}_k$ can be chosen to force $|P^{L(rm)*}_{k,ta}| = |P^{R(rm)}_{ta,k}|$, but not necessarily $P^{L(rm)*}_{k,ta} = P^{R(rm)}_{ta,k}$ since $P^{L(rm)*}_{k,ta}$ and $P^{R(rm)}_{ta,k}$ pick up the same phase under a change of $\mathcal{N}^{(rm)}_k$.
The latter condition must be satisfied for some value of $\mathcal{N}^{(rm)}_k$ if $\ket{y_k^{L(rm)}} = \ket{y_k^{R(rm)}}$ is to hold and thus for $k$ to be an element of the Hermitian subspace.
Thus, checking whether a given Ritz vector $\ket{y_k^{R/L(m)}}$ is an element of the Hermitian subspace can be accomplished by checking the phase of $P^{L(rm)*}_{k,ta}/P^{R(rm)}_{ta,k}$, which will be real and positive for Hermitian vectors.

For a practical prescription, note that for $k$ associated with real Ritz values, the unnormalized Ritz coefficients $\tilde{P}^{R(rm)}_{ta,k} \equiv P^{R(rm)}_{ta,k} / \mathcal{N}^{(rm)}_k$ and $\tilde{P}^{L(rm)}_{k,ta} \equiv \mathcal{N}^{(rm)}_k P^{L(rm)}_{k,ta}$ are related by
\begin{equation}\label{eq:unnorm-ritz-Nk-relation}
    \tilde{P}^{L(rm)}_{k,ta} = |\mathcal{N}^{(rm)}_k|^2 \tilde{P}^{R(rm)*}_{ta,k} ~~~~~ (\lambda^{(rm)}_k \in \mathbb{R}),
\end{equation}
where the the constant of proportionality\footnote{Note that this is an abuse of notation. If there exists some $\mathcal{N}^{(rm)}_k$ that can simultaneously unit-normalize $\ket{y^{R(rm)}_k}$ and $\ket{y^{L(rm)}_k}$, then this constant of proportionality is equal to $|\mathcal{N}^{(rm)}_k|^2$ and manifestly positive. Otherwise, the constant is simply some number whose non-positivity signals states outside the Hermitian subspace; see the discussion around \cref{eq:norm_trick}.} $|\mathcal{N}^{(rm)}_k|^2$ will be real and positive for $k$ in the Hermitian subspace.
Using \cref{eq:unnorm-ritz-Nk-relation}, we see that for real $k$,
\begin{equation}\label{eq:ritz-LR-norm-pos}
\begin{aligned}
    |\mathcal{N}^{(rm)}_k|^2 &= [\bm{P}^{R(rm)\dag} \bm{G} \bm{P}^{R(rm)}]_{kk} ~ ,\\
    |\mathcal{N}^{(rm)}_k|^{-2} &= [\bm{P}^{L(m)} \bm{G} \bm{P}^{L(m)\dag}]_{kk} ~ .
\end{aligned}
\end{equation}
Thus, the Hermitian subspace can be diagnosed as the $k$ for which $\lambda^{(rm)}_k$ is real and both RHSs of \cref{eq:ritz-LR-norm-pos} are real and positive.
For Hermitian $k$, \cref{eq:ritz-LR-norm-pos} also provides the $\mathcal{N}^{(rm)}_k$ to unit-normalize the degenerate left and right Ritz vectors, which are identical to those given by the norm trick, Eq.~\eqref{eq:norm_trick}.

After normalization, the Ritz coefficients may be used to compute overlaps using \cref{eq:Z-factors-ritz-coeffs}, matrix elements using \cref{eq:matrix-elements-PrPl}, and residual bounds using \cref{eq:B-numerator} and \cref{eq:B-denominator} (alternatively, residual bounds can also be computed more directly from Krylov-space quantities; see \cref{app:gevp-Gammas-and-Bs} for a discussion).
This allows computation of all ORR quantities all without reference to the intermediate basis $\aset{v^{R/L}_i}$ and while avoiding the need to compute Krylov coefficients $K^L$ and $K^R$.

Several other observations allow for practical reductions in storage costs.
For example, if the Ritz coefficients have been filtered to the Hermitian subspace and normalized, then $P^{R(rm)}_{k,ta} = P^{L(rm)*}_{ta,k}$, so only one set need be stored.
Under similar conditions, the overlaps (taking the usual convention where they are real) automatically satisfy $Z^{R(m)}_{ka} = Z^{L(m)}_{ka}$.

Notably, rephrasing the problem in terms of GEVP allows natural treatment of the fully oblique case where the right and left vector sets, $\aset{\ket{\xi^R_j}}$ and $\aset{\bra{\xi^L_i}}$, are of different sizes.
In this case, the matrices $G$ and $M$ are rectangular, including only the rows and columns of $\bm{M}$ and $\bm{G}$ corresponding to the chosen Krylov vectors.
The right and left GEVPs, $\bm{G} \bm{P}^R \bm{\Lambda} = \bm{M} \bm{P}^R$ and $\bm{\Lambda} \bm{G} \bm{P}^L = \bm{P}^L \bm{M}$, may be solved to obtain Ritz values and Ritz coefficients as in the square case.
Notably, the matrix Prony method~\cite{Fleming:2023zml} is recovered when one or the other vector set is of length one.
Separately, this may allow application of Rayleigh-Ritz methods to situations where 3-point functions are not computed on a uniform grid, as is standard in many lattice calculations.
Specifically, if $\bm{C}^\mathrm{3pt}(\sigma,\tau)$ is known only for a subset of $\sigma,\tau$, then one can consider RR with appropriately chosen subsets of $\aset{k^{R/L}_{ta}}$ (at the cost of reduced convergence of the Ritz vectors).

\section{Convergence}
\label{sec:convergence}

Oblique Rayleigh-Ritz provides approximations to the eigenvalues and eigenvectors of $T$ using any arbitrary sets of vectors $\aset{\ket{\RRvec^{R/L}_i}}$.
For Hermitian $T$ and $\ket{\RRvec^{R}_i} = \ket{\RRvec^{L}_i}$, the results provide optimal subspace approximations to true eigenvalues and eigenvectors in the max-min sense discussed above.
For non-Hermitian $T^{(d)}$, the same optimality properties are obtained after applying filtering in which non-spurious Ritz vectors are part of the Hermitian subspace.
However, in both cases this does not guarantee that the approximation afforded by a given subspace is ``good'' in other senses.
It is desirable to have some notion of the quality of approximation provided by a given subspace in comparison to others, and in particular one that applies to non-Hermitian cases.
This can be obtained by defining some iteration providing a sequence of ORR estimators that asymptotically converge to true values.

Critically, applications of RR can be grouped into two distinct classes of convergent iterative methods that differ in their convergence rates and other properties, denoted \emph{power iteration} (PI) and \emph{Kaniel-Paige-Saad} (KPS).
Applications of ORR may be similarly grouped into two classes: PI and \emph{oblique}.
As suggested by this terminology, the convergence rate for ORR applications in the power iteration class is independent of whether $T^{(d)}$ is Hermitian or non-Hermitian.
Distinct convergence properties arise between the KPS class of RR applications and the oblique class of ORR applications.
The use of block methods with $r > 1$ leads to faster convergence but does not change the parametric form of the convergence rate and is not denoted separately.

\subsection{PI convergence}

Power-iteration methods improve the quality of ORR approximations by improving the quality of a single vector (or, in the block case, a vector set of fixed size).
The name of this convergence type refers to the power-iteration algorithm, which exploits the fact that applying $T$ to an arbitrary vector will suppress contributions from eigenstates with small eigenvalues.
This can be seen by representing some vector $\ket{\psi}$ in terms of the true eigenvectors $\ket{n}$ of $T$ as $\ket{\psi} = \sum_n \ket{n} \braket{n|\psi}$ and noting
\begin{equation}
    T^t \ket{\psi} = \sum_n (\lambda_n)^t \ket{n} \braket{n|\psi} .
\end{equation}
Under repeated applications of $T$, states with larger eigenvalues will dominate; as $t \rightarrow \infty$, $T^t \ket{\psi}$ converges exponentially quickly to the state with the largest eigenvalue (i.e., the ground state) up to normalization.

The (scalar) power-iteration algorithm can be viewed as applying RR with a single vector, i.e.~using the set $\aset{T^{m-1} \ket{\psi}}$.
In this case, the RR matrices $\bm{G}$ and $\bm{M}$ are each $1 \times 1$, with $G = \braket{\psi| T^{2(m-1)} |\psi}$ and $M = \braket{\psi| T^{2m-1} |\psi}$.
The single Ritz vector $\ket{y_0^{(1,m)}}$, where the $(1,m)$ superscript indicates the $y_0^{(1,m)}$ is the first Ritz vector associated with RR for a particular 1-dimensional subspace labeled by $m$, is thus related trivially to $T^{m-1} \ket{\psi}$ as
\begin{equation}
    \ket{y_0^{(1,m)}} = \frac{ T^{m-1} \ket{\psi} }{ \sqrt{\braket{\psi | T^{2(m-1)} | \psi}} } ~ .
\end{equation} 
The corresponding power-iteration Ritz value $\lambda_0^{(1,m)}$ is obtained as
\begin{equation}
    \lambda_0^{(1,m)} \equiv \braket{y_0^{(1,m)} | T | y_0^{(1,m)}} = \frac{ \braket{\psi | T^{2m-1} | \psi} }{ \braket{\psi | T^{2(m-1)} | \psi} } .
\end{equation}
In application to correlator analyses, the standard effective energy estimator is recovered by defining $E^{(1,m)}_0 \equiv - \ln \lambda^{(1,m)}_0$ and generalizing $(2m-1) \rightarrow t$.
As $m \rightarrow \infty$, $\lambda_0^{(1,m)} \rightarrow \lambda_0$ and $\braket{0 | y_0^{(1,m)}} \rightarrow 1$. 
For large but finite $m$, corrections are suppressed by $\mathcal{O}((\lambda_1/\lambda_0)^{2m}) \sim \mathcal{O}(e^{-2m (E_1 - E_0)})$~\cite{Parlett}.

Generalizing to oblique power iteration (OPI), as necessary to treat non-Hermitian $T \neq T^\dagger$ and/or $\ket{\psi^R} \neq \ket{\psi^L}$, is straightforward.
OPI corresponds to ORR with one right vector and one left vector, i.e.~with right and left vector sets $\aset{T^{m-1} \ket{\psi^R}}$ and $\aset{\bra{\psi^L} T^{m-1}}$.
The R/L Ritz vectors $\ket{y^{R/L(1,m)}_0}$ are obtained similarly trivially as in the non-oblique case, so the single OPI Ritz value is
\begin{equation}
    \lambda^{(1,m)}_0 
    \equiv \braket{y^{L(1,m)}_0 | T | y^{R(1,m)}}
    = \frac{ 
        \braket{\psi^{L} | T^{2m-1} | \psi^{R}}
    }{
        \braket{\psi^{L} | T^{2(m-1)} | \psi^{R}}
    } ~ .
\end{equation}
The expression is unchanged from the non-oblique case except for the addition of $R/L$ labels and, in application to correlator analyses, prescribes an identical effective energy estimator.
As $m \rightarrow \infty$, $\lambda^{(1,m)}_0$ converges to the eigenvalue with the largest absolute magnitude, and $\ket{y^{R/L(1,m)}}$ to its corresponding R/L eigenvectors.
Corrections are again suppressed by $\mathcal{O}((\lambda_1/\lambda_0)^{2m})$, so the convergence rate is not affected by the generalization to the oblique case.

The block generalization of oblique power iteration can be thought of as ORR using right and left sets $\aset{T^{m-1} \ket{\psi^R_a}}$ and $\aset{\bra{\psi^L_a} T^{m-1}}$, each of $r$ vectors indexed by $a$.
The vectors within each set are assumed to be linearly independent, but the left and right sets may share some vectors or be identical.
The ORR matrices are then
\begin{equation}
\begin{split}
    G_{ab} &= \braket{\psi^L_a | T^{2(m-1)} | \psi^R_b}, \\
    M_{ab} &= \braket{\psi^L_a | T^{2m-1} | \psi^R_b} .
    \end{split}
\end{equation}
Taking the oblique convention $\bm{\RRmat}^L = \bm{G}$ and $\bm{\RRmat}^R = \bm{1}$ for simplicity, RR prescribes obtaining the block power-iteration Ritz values as the eigenvalues of $\bm{G}^{-1} \bm{M}$.
We can immediately see the connection to GEVP methods by noting these Ritz values, denoted $\lambda_k^{(r,m)}$, are equivalent to the generalized eigenvalues found by the GEVP
\begin{equation}
\label{eq:gevp-def}
    \bm{G} \vec{g}_k^{(r,m)} = \bm{M} \vec{g}_k^{(r,m)} \lambda_k^{(r,m)}.
\end{equation}
For large but finite $m$, corrections to rank-$r$ GEVP are suppressed by $\mathcal{O}((\lambda_r/\lambda_0)^{2m}) \sim \mathcal{O}(e^{-2m (E_r - E_0)})$~\cite{Luscher:1990ck,Blossier:2009kd}.
Iterative methods achieving this convergence rate for a rank-$r$ subspace, for which $r=1$ corresponds to the power-iteration convergence above, are designated in general as achieving PI convergence below.

\subsection{KPS convergence}

Lanczos methods use a distinct iterative construction to achieve improved convergence properties, specifically applying RR to a sequence of (block) Krylov subspaces with increasing dimension $rm$.
This has long been known to provide superior convergence to power-iteration methods~\cite{Parlett}.
The convergence of (block) KRR methods is quantified by Kaniel-Paige-Saad convergence theory~\cite{Kaniel:1966,Paige:1971,Saad:1980}.
In particular, for the case of small gaps where convergence is most challenging, the KPS bound shows that differences between Ritz values and $T$ eigenvalues are suppressed by $\mathcal{O}(e^{-4m\sqrt{E_r-E_0}})$ and are thus exponentially suppressed compared to the corresponding PI differences.

The general form of the KPS bound is~\cite{Saad:1980}
\begin{equation}\label{eq:KPS}
    0 \leq \frac{ \lambda_n - \lambda_n^{(rm)} }{ \lambda_n - \lambda_{\infty} } \leq \left[ \frac{ K_n^{(rm)} \tan \theta_n}{ T_{m-n-1}(\Gamma_n^r)} \right]^2,
\end{equation}
where the $T_k(x)$ are Chebyshev polynomials of the first kind defined by $T_k(\cos x) = \cos(k x)$,
\begin{equation}
\begin{split}
    \Gamma_n^r &\equiv 1 + \frac{2(\lambda_n - \lambda_{n+r})}{\lambda_{n+r} - \lambda_{\infty}} = 2 e^{E_{n+r}-E_n} - 1,
    \end{split}
    \label{eq:kps-gamma}
\end{equation}
and
\begin{equation}
    K_n^{(rm)} \equiv \prod_{l=0}^{n-1} \frac{\lambda_l^{(rm)} - \lambda_{\infty}}{\lambda_l^{(rm)} - \lambda_n}, \hspace{20pt} n > 0.
    \label{eq:kps-K}
\end{equation}
Here, $K_0^{(m)} \equiv 1$ and $\lambda_{\infty}$ is the smallest eigenvalue of $T$, which for a bounded infinite-dimensional operator $T$ must be $\lambda_{\infty} = 0$.
The angle $\theta_n$ in \cref{eq:KPS} is the angle between the subspace $\mathcal{K}^{(r)}$ spanned by the initial subspace,
\begin{equation}
    \tan^2\theta_n = ||\ket{n} - \ket{\hat{x}_n}||^2.
\end{equation}
The explicit construction of this vector and further details of computing the block KPS bound are discussed in Ref.~\cite{Hackett:2024nbe}.
For the ground state, the KPS bound simplifies to 
\begin{equation}\label{eq:KPS0} \begin{split}
    0 \leq \frac{\lambda_0 - \lambda_0^{(m)}}{\lambda_0} &\leq  \frac{\tan^2\theta_0}{T_{m-1}(2 e^{\delta_r} - 1)^2},
    \end{split}
\end{equation}
where $\delta_r = E_r - E_0$ and 
\begin{equation}
    \tan^2\theta_0 = \left[ \bm{Z} \bm{C}(0)^{-1} \bm{Z}^\dagger \right]^{-1}_{00} - 1,
\end{equation}
where $\bm{Z}$ is the $r \times r$ matrix defined by $Z_{na} = \braket{n | \psi_a}$ with $n \in \{0,\ldots,r-1\}$.
The asymptotic behavior simplifies as
\begin{equation}\label{eq:block_KPS}
    0 \leq \frac{\lambda_0 - \lambda_0^{(rm)}}{\lambda_0 } \lesssim  4 \tan^2\theta_0  \times \begin{cases} e^{-2 (m-1) \delta_r }    &  \delta_r \gg 1 \\
    e^{-4(m-1)\sqrt{\delta_r}} &  \delta_r \ll 1 
    \end{cases} .
\end{equation}
using $T_{m}(x) \approx \frac{1}{2}(x + \sqrt{x^2 - 1})^{m}$.
The $e^{-4(m-1)\sqrt{\delta_r}}$ factor arising from the expansion of $T_{m-1}(2 e^{\delta_r} - 1)^{-2}$ for small $\delta_r$ is the source of faster convergence for RR methods in comparison with power-iteration methods, where the KPS bound is applicable.
For large $\delta_r$ both power-iteration and KPS methods converge rapidly.

The advantages of KRR methods over power iteration are unsurprising when viewed from the lens that both are applications of RR with different choices of vector sets:
Rayleigh-Ritz with Krylov spaces use a superset of the information available to power iteration.
This predictably leads to improved quality of approximation of both eigenvalues and eigenvectors.
Moreover, using Krylov spaces as the vector sets provides a qualitative gain of capabilities over power iteration: as $m \rightarrow \infty$, the full eigensystem of $T$ is recovered.

Use of power iteration and Krylov-space iterations are not mutually exclusive.
Applying $T$ to all elements of a Krylov-space vector set provides a second set of vectors with improved overlap with the low-lying states.
In correlator analyses, this corresponds to applications of methods like Lanczos and Prony to a ``sliding window'' of the correlator.
This was the strategy used in early lattice QCD applications of Prony's method~\cite{Fleming:2004hs,Lin:2007iq,Fleming:2009wb} and GPOF~\cite{Aubin:2010jc,Aubin:2011zz,Green:2014xba,Schiel:2015kwa,Ottnad:2017mzd,Fischer:2020bgv} in which spurious-state filtering was not applied.
Such a sliding-window formulation may be useful when it is computationally unfeasible to perform KRR on a set of correlator matrices, but applying KRR methods to the full set of correlator matrices is still guaranteed to achieve faster convergence.

\subsection{Oblique convergence}

The convergence of KRR methods in the oblique case, i.e.~of OKRR and OBKRR, is more complicated~\cite{Saad:1981,Saad:1982,Saad:2011}.
The factor of $1/T_{m-n-1}(\Gamma_n^r)$ appearing in the ground-state KPS bound for $r=1$ leading to rapid exponential convergence for Hermitian $T$ can be understood as arising from a more general bound on the accuracy of Krylov-subspace projection~\cite{Saad:1980}.
After $m$ Lanczos iterations, the accuracy with which $\ket{n}$ can be reconstructed within the subspace associated with $\Pm{rm}$ can be quantified by the size of 
\begin{equation}
    \Pi^{(rm)}_n \equiv \frac{ ||(1 - \Pm{rm}) \ket{n}|| }{ || \Pm{rm} \ket{n} || },
\end{equation}
which vanishes if $\ket{n}$ is reconstructed perfectly and thus $\Pm{rm} \ket{n} = \ket{n}$, in which case some Ritz vector has perfectly converged to $\ket{n}$. 
This quantity can be proven to decrease monotonically from its $m=1$ value and satisfies the bound 
\begin{equation}\label{eq:eval_symmetric}
    \Pi^{(rm)}_n \leq \frac{K_n^{(rm)}}{T_{m-n-1}(\Gamma_n^r)} \Pi^{(r)}_n.
\end{equation}
This bound underlies the KPS bound and rapid convergence of Ritz values/vectors for Hermitian $T$.

To make contact with the oblique case for non-Hermitian $T$, it is helpful to first consider the closely related problem of solving a system of linear equations using iterative Krylov subspace methods~\cite{Saad:1981,Saad:1982,Saad:2011}.
This could arise for instance in the simpler problem of solving the linear system of equations
\begin{equation}\label{eq:linear}
\begin{split}
    \sum_j  T_{ij} x_{j} = b_i,
\end{split}
\end{equation}
for $\vec{x}$ given $T_{ij}$ and $\vec{b}$.
If $T_{ij}$ is Hermitian, then the conjugate-gradient analog of symmetric Lanczos provides convergent approximations whose finite-iteration errors are bounded by an expression analogous to \cref{eq:eval_symmetric},~\cite{Saad:1981}
\begin{equation}\label{eq:solver}
    ||(1 - \Pm{\ell}) \vec{x}|| \leq \frac{1}{T_{\ell}(\gamma)},
\end{equation}
where $\ell$ is the iteration count, $\mathcal{P}^{(\ell)}$ is a Krylov-subspace projector, and $\gamma = (\eta_{\rm max} + \eta_{\rm min})/(\eta_{\rm max} - \eta_{\rm min})$ is defined in terms of the minimum and maximum eigenvalues of the Hermitian matrix $T_{ij}$ appearing in the linear system.
If instead $T_{ij}$ is non-Hermitian, then oblique Lanczos biorthogonalization can be used to obtain a solution to Eq.~\eqref{eq:linear} whose finite-iteration errors are bounded by the more complicated expression~\cite{Saad:1981}
\begin{equation}\label{eq:oblique_KPS}
    ||(1 - \Pm{\ell}) \vec{x}|| \leq \left[ \sum_{j=1}^{\ell+1} \prod_{l=1, l\neq j}^{\ell+1} \frac{|\eta_l| }{|\eta_j - \eta_l|} \right]^{-1}.
\end{equation}
Here, the correct ordering of the complex eigenvalues $\eta_l$ of $T_{ij}$ appearing in the product is not known constructively;  Ref.~\cite{Saad:1981} shows that there exists some ordering for which \cref{eq:oblique_KPS} is valid.
For some spectra, \cref{eq:oblique_KPS} leads to exponential convergence with rates on the order of the gaps in the spectrum as demonstrated in Refs.~\cite{Saad:1981,Saad:1982}.
However, the convergence rate depends sensitively on the distributions of $T_{ij}$ eigenvalues in the complex plane and can be much slower than the convergence rate of the Hermitian case governed by Eq.~\eqref{eq:solver} in some cases~\cite{Saad:1981,Saad:2011}.
In particular, if the spectrum of $T_{ij}$ can be bounded to lie within an ellipse centered at $c \in \mathbb{R}$ with foci $c \pm e$ and semi-major axis $a$ then \cref{eq:oblique_KPS} can be replaced with the explicit bound~\cite{Saad:1981}
\begin{equation}
    ||(1 - \Pm{\ell}) \vec{x} || \leq \frac{T_{\ell}(a/e)}{|T_{\ell}(c/e)|}.
\end{equation}
For a spectrum lying close to the real axis, corresponding to small $a$, the numerator approaches $T_{\ell}(1) = 1$ and the denominator approaches $T_{\ell}(\gamma)$, indicating that the oblique Lanczos convergence-rate bound coincides with that of the symmetric case.
However, the oblique Lanczos convergence rate decreases as the eccentricity decreases and the spectrum shifts from an oval around the real line towards a circle in the complex plane.

These bounds on the convergence of linear solves using symmetric and oblique Lanczos provide qualitative but not necessarily quantitative guidance on the convergence of OKRR for eigenvalue approximation.
The general convergence theory of OKRR is complicated by the fact that in some pathological cases where $\mathcal{K}^{R(m)}$ and $\mathcal{K}^{L(m)}$ are orthogonal the algorithm suffers from an ``incurable breakdown'' and terminates~\cite{Parlett:1985,Nachtigal:1993,Saad:2011}.
Such a breakdown is unlikely to occur in an exact sense for noisy correlator data.
However, small but non-zero overlap between the $\ket{\psi^R_a}$ and  $\ket{\psi^L_a}$  can sometimes lead to pathologically slow convergence behavior called ``stagnation'', which may be a practical concern~\cite{Leyk:1997,Gaaf:2016,Saad:2011}.

\subsection{Filtered convergence}
\label{sec:filtering-convergence}

As already discussed in \cref{sec:filtering}, in the diagonal case where $\ket{\psi^R_a} = \ket{\psi^L_a}$, it is possible to identify a Hermitian subspace $\herm$ of states with real Ritz values and degenerate right/left Ritz vectors, i.e., $\lambda^{(rm)}_k \in \mathbb{R}$ and $\ket{y^{R(rm)}_k} = \ket{y^{L(rm)}_k}$ for all $k \in h$.
In practice, states in the Hermitian subspace may be identified using the tests defined in \cref{sec:rr-practice}.
As noted in \cref{sec:filtering}, ORR results filtered to the Hermitian subspace inherit the optimality guarantees of non-oblique RR.

States in the Hermitian subspace similarly inherit KPS convergence for iterations for $m \leq m_{\herm}$, where $m_{\herm}$ is defined as the largest iteration for which all Ritz vectors are in the Hermitian subspace, or in other words the largest $m$ for which $d_{\herm} \equiv \dim(\herm) = rm$.
This is already sufficient to demonstrate that FRR converges exponentially faster than power-iteration methods on noisy correlator (matrix) data for $m \leq m_{\herm}$. 

For $m \geq m_{\herm}$, analysis must be viewed as ORR and the KPS bound does not apply.
Convergence for $m > m_{\herm}$ continues to occur, but only in the (hard-to-quantify) sense of oblique Lanczos described above.
The KPS bound for $m = m_{\herm}$ is therefore the most conservative bound that can be placed on the convergence of FRR for $m \geq m_{\herm}$,
\begin{equation}\label{eq:KPS_FRR} \begin{split}
    0 \leq \frac{\lambda_0 - \lambda_0^{(rm)}}{\lambda_0} &\leq  \frac{\tan^2\theta_0}{T_{m_{\herm}-1}(2 e^{\delta_r} - 1)^2} \\
    &\lesssim  4 \tan^2\theta_0  \times \begin{cases} e^{-2 (m_{\herm}-1) \delta_r }  
    &  \delta_r \gg 1 \\
    e^{-4(m_{\herm}-1)\sqrt{\delta_r}} &  \delta_r \ll 1 
    \end{cases}
    ,
    \end{split}
\end{equation}

It is possible for the dimension of the Hermitian subspace, $d_{\herm}$, to continue growing for $m \geq m_{\herm}$ in applications to noisy correlators even though it is no longer guaranteed to grow by $r$ at each iteration.
This means that application of FRR can still be viewed as an application of Rayleigh-Ritz to successively larger subspaces in which $P_{\herm} T^{(rm)} P_{\herm}$ acts as a Hermitian operator.
The Ritz values provide optimal approximations to the $d_{\herm}$ eigenvalues of $P_{\herm} T^{(rm)} P_{\herm}$, and it is therefore reasonable to conjecture that the approximation quality of the Ritz values is controlled by $d_{\herm}$ even for $m > m_{\herm}$. 
We therefore conjecture that in the scalar case
\begin{equation}\label{eq:KPS_FRR_conj} \begin{split}
    0 \leq \frac{\lambda_0 - \lambda_0^{(m)}}{\lambda_0} &\stackrel{?}{\leq}  \frac{\tan^2\theta_0}{T_{d_{\mathcal{H}}-1}(2 e^{\delta_1} - 1)^2} \\
    &\lesssim  4 \tan^2\theta_0  \times \begin{cases} e^{-2 (d_{\mathcal{H}}-1) \delta_1 }  
    &  \delta_1 \gg 1 \\
    e^{-4(d_{\mathcal{H}}-1)\sqrt{\delta_1}} &  \delta_1 \ll 1 
    \end{cases}
    .
    \end{split}
\end{equation}
It is not obvious how to extend Eq.~\eqref{eq:KPS_FRR_conj} to the block case, in particular since $d_{\mathcal{H}}$ is not constrained to be a multiple of $r$ for $r \geq 2$.

For asymmetric correlators, $d_{\mathcal{H}} = 0$ even in the infinite-statistics limit because $\ket{\psi_a^R} \neq \ket{\psi_a^L}$ implies $\ket{y_k^R} \neq \ket{y_k^L}$.
Therefore FRR applied to asymmetric correlators never leads to KRR convergence, only convergence in the oblique Lanczos sense.

\section{Prony \& the Unique Exponential Decomposition}
\label{sec:prony-uniqueness}

Equivalences between oblique block Lanczos, GEVP, and  GPOF as framed in the lattice literature~\cite{Aubin:2010jc,Aubin:2011zz} are clear when they are all viewed as applications of ORR to a transfer matrix acting in Hilbert space, as discussed above.
It is less obvious why these methods should be equivalent to  Prony and GPOF as framed in the mathematics literature~\cite{Hua:1989,Sarkar:1995}, which are motivated purely algebraically.
One way of understanding this equivalence is by viewing all of these methods as members of a broader equivalence class that provides unique eigenvalue and eigenvector approximations.

This section proves the uniqueness of all (block) correlator analysis methods that solve a finite-dimensional spectrum exactly given sufficiently many input correlator values.
The proof essentially follows the derivations of Prony's method \cite{Prony} and its block generalization \cite{Fleming:2023zml}, which directly consider the algebraic relation between the correlator data and the parameters in the decomposition.

More precisely, we assume that we have at least one decomposition of the form $\bm{C}(t) = \sum_{k=0}^{m-1} \bm{a}_k \lambda_k^t$ where the $\lambda_k$ are non-degenerate and the $\bm{a}_k$ are rank one.
Uniqueness of this correlator decomposition holds for scalar correlators whenever the Hankel matrix $H^{(0)}$ is invertible.
Uniqueness holds for correlator matrices whenever $\bm{H}^{(0)}$ is invertible and a closely related generalized Vandermonde matrix is full rank.
It is improbable for these conditions to be exactly violated for noisy data, and in explicit examples it is straightforward to compute the ranks of these Hankel / generalized Vandermonde matrices and check whether these conditions hold.

\subsection{Prony's method \& uniqueness in the scalar case}

To begin, suppose we have some set of $m$ many distinct $\aset{\lambda_k}$ indexed by $k \in \{0,\ldots,m-1\}$.
These define a \emph{characteristic polynomial},
\begin{equation}\label{eq:char_poly}
    p^{(m)}(\lambda) 
    \equiv \prod_{k=0}^{m-1} (\lambda - \lambda_k) 
    \equiv \lambda^m - \sum_{s=0}^{m-1} \gamma_s \lambda^s ~ .
\end{equation}
whose roots are the $\aset{\lambda_k}$ by construction, i.e., $p^{(m)}(\lambda_k) = 0$ for any $\lambda_k$.
In the second definition, the $m$ polynomial coefficients or \emph{linear prediction coefficients}, $\aset{\gamma_s}$, are defined in terms of $\lambda_k$.
They may be computed simply by expanding $\prod_k (\lambda - \lambda_k)$ and grouping terms by powers of $\lambda$, so $\gamma_0 = \prod_k (-\lambda_k)$, $\gamma_1 = \sum_k \prod_{l \neq k} (-\lambda_l)$, etc.
Because $p^{(m)}(\lambda_k) = 0$ for any $\lambda_k$, a simple rearrangement yields the \emph{recurrence relation}, $\lambda_k^m = \sum_{s=0}^{m-1} \gamma_s \lambda_k^s$, which holds for all $\lambda_k$. Multiplying both sides of this recurrence relation by $\lambda_k^{t_0}$ gives 
\begin{equation}\label{eq:prony-scalar-recurrence}
    \lambda_k^{m+t_0} = \sum_{s=0}^{m-1} \gamma_s \lambda_k^{s+t_0}.
\end{equation}
Because the equation holds for all $\lambda_k$ with the same shared values of $\gamma_s$, we may take an arbitrary linear combination,
\begin{equation}\label{eq:prony-scalar-decomp-recurrence}
    \sum_{k=0}^{m-1} a_k \lambda_k^{m+t_0} = \sum_{k=0}^{m-1} \sum_{s=0}^{m-1} a_k \gamma_s \lambda_k^{s+t_0} ~ ,
\end{equation}
which holds for any set of $m$ values $\aset{a_k}$.

Now, suppose we have some set of $2m$ data points $\aset{C(t)}$ indexed by $t \in \{0,\ldots,2m-1\}$.
We assume they admit at least one decomposition of the form $C(t) = \sum_{k=0}^{m-1} a_k \lambda_k^t$.
The objective is now to prove that they admit \emph{only one} such decomposition, i.e., that the $2m$ parameters in such a decomposition, $a_k$ and $\lambda_k$, are uniquely defined given a fixed set of $2m$ values of $C(t)$.

The necessary proof amounts to following the steps to apply Prony's method.
This starts with determining the $\gamma_s$.
We have assumed a decomposition exists, so by \cref{eq:prony-scalar-decomp-recurrence}, it follows that
\begin{equation}\label{eq:scalar-prony-corr-recurrence}
    C(m+t_0) = \sum_{s=0}^{m-1} C(t_0+s) \gamma_s,
\end{equation}
for all $t_0 \in \{0,\ldots,m-1\}$ simultaneously.
This amounts to a system of $m$ linear constraints on the $m$ unknowns $\gamma_s$, defined by the $m \times m$ Hankel matrix $H^{(0)}_{t_0,s} = C(t_0+s)$ for $t_0,s \in \{0,\ldots,m-1\}$.
If $H^{(0)}$ is invertible, the system admits a unique solution and $\gamma_s$ is uniquely defined from $C(t)$.
Once the $\gamma_s$ are obtained, we can construct $p^{(m)}$ and find its $m$ roots, $\lambda_k$, assumed not to be degenerate.
Therefore, the $\lambda_k$ are uniquely defined from $\gamma_s$, and thus from $C(t)$.

What remains is to show the amplitudes $a_k$ are also uniquely defined from $C(t)$.
To obtain them, observe that if the $\lambda_k$ are known, the expression $C(t) = \sum_k a_k \lambda_k^t$, which holds for all $t \in \{0,\ldots,2m-1\}$ simultaneously, can be viewed as $2m$ linear constraints on the $m$ unknowns $a_k$.
This may be expressed more precisely in terms of the $2m \times m$ \emph{Vandermonde matrix}, $V_{tk} \equiv \lambda_k^t$ for $k \in \{0,\ldots,m-1\}$ and $t \in \{0,\ldots,2m-1\}$, i.e.,
\begin{equation}\label{eq:vandermonde-matrix}
    V \equiv \begin{bmatrix}
        1 & 1 & \cdots & 1 \\
        \lambda_0   & \lambda_1   & \cdots   & \lambda_{m-1} \\
        \lambda_0^2   & \lambda_1^2   & \cdots   & \lambda_{m-1}^2 \\
        \vdots      & \vdots      & \ddots   & \vdots \\
        \lambda_0^{2m-1} & \lambda_1^{2m-1} & \cdots & \lambda_{m-1}^{2m-1}
    \end{bmatrix} ~ .
\end{equation}
The amplitudes are thus constrained by the overdetermined system of equations
\begin{equation}
\label{eq:scalar-vandermonde-system}
    C = V a ~ ,
\end{equation}
written in matrix-vector notation with $C$ and $a$ as column vectors of length $2m$ and $m$, respectively.
Recall that we have assumed at least one solution for $a$ exists; to assess whether multiple are possible, suppose two distinct solutions, $a$ and $a'$.
Because $C = V a$ and $C = V a'$, it follows that $V(a-a') = 0$.
This implies that $a = a' + \zeta$, where $\zeta$ is any vector in the (right) null space of $V$, i.e.~for which $V \zeta = 0$.
However, as long as all $\lambda_k$ are distinct (as already assumed), the columns of $V$ are all linearly independent, and thus the rank of $V$ is exactly $m$---i.e., $V$ has ``full column rank''.
This means that $V$ has no nontrivial (right) null space, such that only $\zeta=0$ satisfies $V \zeta = 0$.
We thus conclude that $a = a'$, and there is a unique solution for the amplitudes $a_k$ given $C(t)$ and the $\lambda_k$ already (uniquely) determined.
This completes the proof that the correlator decomposition is unique.

\subsection{Block Prony \& uniqueness in the block case}

The corresponding starting point in the block case is that an arbitrary sequence of $2m$ $r \times r$ complex matrices, $C_{ab}(t)$ for $t \in \{0,\ldots,2m-1\}$, admits a decomposition of the form
\begin{equation}
    C_{ab}(t) = \sum_{k=0}^{rm-1} a_{k,ab} \lambda_k^t,
\end{equation}
where the amplitudes $\bm{a}_k$ are rank one, i.e., can be parametrized as an outer product
\begin{equation}
    a_{k,ab} = Z^{R*}_{ka} Z^L_{kb} ~ .
\end{equation}
Each of $\vec{Z}^R_k$ and $\vec{Z}^L_k$ are defined only up to an overall complex constant, but this ambiguity cancels in the product.
Similar to the scalar case, the proof that this decomposition is unique follows the steps of the block generalization of Prony's method, as introduced in Ref.~\cite{Fleming:2023zml}.
However, a slightly different treatment is required to fully exploit the matrix structure and obtain a decomposition of the desired form.
To see this, note that in principle, as in the scalar case, each term $\lambda_k$ corresponds to the roots of a (scalar) characteristic polynomial $\lambda^m - \sum_{s=0}^{rm-1} \gamma_s \lambda^s$.
However, the corresponding recurrence is of depth $rm$ while the correlator matrix is of length $2m$ only, which means that while the relation 
\begin{equation}
    \sum_k \bm{a_k} \lambda_k^{rm} = \sum_k \sum_{s=0}^{rm-1} \bm{a_k} \lambda_k^s,
\end{equation}
holds formally, it does not provide any useful relation between correlator data analogous to \cref{eq:prony-scalar-recurrence}: the LHS corresponds to $\bm{C}(t)$ evaluated at $t = rm$, but $\bm{C}(t)$ is only defined for $t \leq 2m-1$. 

Instead, block Prony proceeds by considering a block generalization of the characteristic polynomial.
The additional matrix structure allows for distinct right- and left-acting generalizations
\begin{equation}\label{eq:block-char-poly}
    \bm{p}^{R/L(m)}(\lambda) \equiv \lambda^{m} \bm{1} - \sum_{\sigma=0}^{m-1} \bm{\Gamma}^{R/L}_\sigma \lambda^\sigma ~ ,
\end{equation}
each of which will lead to the same decomposition.
In brief, the linear prediction matrices, $\bm{\Gamma}^{R/L}_\sigma$, will be defined so that the matrix-valued polynomials $\bm{p}^{R/L(m)}(\lambda)$ satisfy
\begin{equation}\label{eq:block-prony-roots-proof-1}
    \bm{a}_k \bm{p}^{R(m)}(\lambda_k) = 0
    ~~\text{ and }~~
    \bm{p}^{L(m)}(\lambda_k) \bm{a}_k = 0,
\end{equation}
for all $k \in \{0,\ldots,rm-1\}$, which in turn will ensure that the $\lambda_k$ arise as the $rm$ roots of either scalar polynomial $\det \bm{p}^{R(m)}(\lambda)$ or $\det \bm{p}^{L(m)}(\lambda)$.\footnote{\label{foot:LR-lambda-caveat}That the roots of these two polynomials coincide is nontrivial. Proof is deferred until block companion matrices are introduced below, from which perspective it is obvious.}

The block analog to \cref{eq:prony-scalar-recurrence} is obtained by combining \cref{eq:block-char-poly,eq:block-prony-roots-proof-1} to give
\begin{equation}
\begin{aligned}
    \bm{a}_k \lambda_k^{m} &= \bm{a}_k \sum_{\sigma=0}^{m-1} \bm{\Gamma}^R_\sigma \lambda_k^\sigma ~ , \\
    \lambda_k^{m} \bm{a}_k &= \sum_{\sigma=0}^{m-1} \bm{\Gamma}^L_\sigma \lambda_k^\sigma \bm{a}_k ~ .
\end{aligned}
\end{equation}
Multiplying both sides by $\lambda_k^{t_0}$, summing over $k$, and recognizing $\bm{C}(t)$ then gives the block generalizations of \cref{eq:scalar-prony-corr-recurrence},
\begin{equation}
\begin{aligned}
    \bm{C}(m+t_0) 
    &= \sum_{\sigma=0}^{m-1} \bm{C}(\sigma + t_0) \bm{\Gamma}^R_\sigma ~ ,
    \\
    &= \sum_{\sigma=0}^{m-1} \bm{\Gamma}^L_\sigma \bm{C}(\sigma + t_0) ~ .
\end{aligned}
\end{equation}
Demanding these hold simultaneously for $t_0 \in \{0,\ldots,m-1\}$ defines block systems of linear equations
\begin{equation}
\begin{bmatrix}
    \bm{C}(m) \\
    \bm{C}(m+1) \\
    \vdots \\
    \bm{C}(2m-1)
\end{bmatrix}
= 
\bm{H}^{(0)} 
\begin{bmatrix}
    \bm{\Gamma}^R_0 \\
    \bm{\Gamma}^R_1 \\
    \vdots \\
    \bm{\Gamma}^R_{m-1} \\
\end{bmatrix},
\end{equation}
and
\begin{equation}
\begin{bmatrix}
    \bm{C}(m) &
    \cdots &
    \bm{C}(2m-1)
\end{bmatrix}
= 
\begin{bmatrix}
    \bm{\Gamma}^L_0 &
    \cdots
    \bm{\Gamma}^L_{m-1}
\end{bmatrix}
\bm{H}^{(0)} ,
\end{equation}
in terms of the $(m,r) \times (m,r)$ block Hankel matrix $\bm{H}^{(0)}$.
Each system amounts to $mr^2$ constraints on $mr^2$ unknowns.
Thus, if $\bm{H}^{(0)}$ is invertible, each system admits a unique solution for the $\bm{\Gamma}^{R/L}_\sigma$.
With the $\bm{\Gamma}^{R/L}_{\sigma}$ uniquely defined by the correlator matrix through this solution, the $\bm{p}^{R/L(m)}(\lambda)$ are uniquely defined by \cref{eq:block-char-poly} and guaranteed to satisfy \cref{eq:block-prony-roots-proof-1}.

Uniqueness of the linear prediction matrices can be used to show that  the $\lambda_k$ are uniquely specified as the roots of  $\bm{p}^{R/L(m)}(\lambda)$ as follows.
If $\bm{p}^{R/L(m)}(\lambda_k)$ were invertible, then \cref{eq:block-prony-roots-proof-1} would imply that $\bm{a}_k = 0$, but the $\bm{a}_k$ are assumed to be nonzero.
This contradiction implies that $\bm{p}^{R/L(m)}(\lambda_k)$ must be non-invertible, i.e., it must have at least one zero eigenvalue.
It follows that
\begin{equation}\label{eq:block-char-poly-det}
    \det \bm{p}^{R/L(m)}(\lambda_k) = 0,
\end{equation}
for all $rm$ different $\lambda_k$.
This exhaustively determines the roots of the order $rm$ polynomials $\det \bm{p}^{R/L(m)}(\lambda)$: the only zeros in $\lambda$ are the $\lambda_k$.
It follows that the $\lambda_k$ are uniquely specified by the $\bm{\Gamma}^{R/L}_{\sigma}$ and therefore by the input correlator matrix.
Note that either the $\bm{\Gamma}^{R}_{\sigma}$ or $\bm{\Gamma}^{L}_{\sigma}$ is sufficient to specify the $\lambda_k$; the fact that both lead to the same $\lambda_k$ is proven in \cref{sec:companion} below.
This exhaustion furthermore implies, along with the assumption that $\lambda_k$ are distinct, that there is a single, non-degenerate zero eigenvalue of $\bm{p}^{R/L(m)}(\lambda_k)$ at each $\lambda_k$, as otherwise $\det \bm{p}^{R/L(m)}(\lambda)$ would have repeated roots.

Unlike the equivalent quantities in the scalar limit, the $\bm{\Gamma}^{R/L}_\sigma$ also contain information about the amplitudes $\bm{a}_k$.
Note that we have assumed the $\bm{a}_k$ are rank one, and thus may be parametrized as an outer product,
\begin{equation}
    \bm{a}_k \equiv \vec{z}^L_k \vec{z}^{R\dagger}_k ~ .
\end{equation}
Each of $\vec{z}^R_k$ and $\vec{z}^L_k$ are each defined only up to an overall complex constant, but this ambiguity cancels in the product.
Inserting this parametrization into \cref{eq:block-prony-roots-proof-1} yields
\begin{equation}\label{eq:block-prony-amp-eigeneq}
    \vec{z}^{R\dagger}_k \bm{p}^{R(m)}(\lambda_k) = 0
    ~~\text{ and }~~
    \bm{p}^{L(m)}(\lambda_k) \vec{z}^L_k = 0 ~ ,
\end{equation}
where in each case the other vector has been removed (by the assumption that the $\bm{a}_k$ are rank one, each vector must be non-zero).
Now, recall that the zero eigenvalue of $\bm{p}^{R/L(m)}(\lambda_k)$ at each $\lambda_k$ is non-degenerate.
This means that \Cref{eq:block-prony-amp-eigeneq} implies $\vec{z}^{R\dagger}_k$ is the left eigenvector of $\bm{p}^{R(m)}(\lambda_k)$ associated with its zero eigenvalue, and $\vec{z}^{L}_k$ is the right eigenvector of $\bm{p}^{L(m)}(\lambda_k)$ associated with its zero eigenvalue.
However, this is insufficient to fully constrain the amplitudes: eigenvector normalization is a matter of convention, so\footnote{While the normalization of left eigenvectors can be defined consistently with the right ones, the left and right eigenvectors here are taken from different matrices, so the ambiguity is unavoidable.} this determines the amplitudes $\bm{a}_k$ only up to an overall complex constant $\alpha_k$ as
\begin{equation}\label{eq:block-prony-unnormed-amplitude}
    \bm{a}_k = \alpha_k \tilde{\bm{a}}_k,
\end{equation}
where the unnormalized amplitude $\tilde{\bm{a}}_k$ is the appropriate outer product of eigenvectors of $\bm{p}^{R/L(m)}(\lambda_k)$ with whatever choice of normalization convention.
Up to this normalization ambiguity, the $\tilde{\bm{a}}_k$ follow uniquely from the  $\bm{p}^{R/L(m)}(\lambda_k)$ and therefore from the input correlator matrix.

What remains is to show that the amplitude normalizations $\alpha_k$ are uniquely determined.
Note that in the scalar limit $r=1$, the eigenvectors of $\bm{p}^{R/L(m)}(\lambda_k)$ are trivial and we recover $\alpha_k = a_k$.
We thus see that what is required to complete the block case is to generalize the scalar Vandermonde matrix procedure.
In the block case,
\begin{equation}
    C_{ab}(t) = \sum_k \tilde{a}_{k,ab} \lambda_k^t \alpha_k ,
\end{equation}
holds simultaneously for all $t \in \{0,\ldots,2m-1\}$ and $a,b \in \{0,\ldots,r-1\}$, for a total of $2mr^2$ linear constraints on the $mr$ unknowns $\alpha_k$.
We may define a block generalization of the Vandermonde matrix,
\begin{equation}
    \mathcal{V}_{tab,k} \equiv \tilde{a}_{k,ab} \lambda_k^t = \tilde{a}_{k,ab} V_{tk} ~ ,
\end{equation}
which, viewing $tab$ as a composite index, can be considered a $2mr^2 \times mr$ matrix.
The second equality emphasizes that this is simply a scalar Vandermonde matrix (in this case, $2m \times mr$) augmented by the block structure of the unnormalized amplitudes $\tilde{\bm{a}}_k$.
Viewing $\alpha_k$ as a length-$mr$ vector and $C_{ab}(t)$ as a length-$2mr^2$ vector in the composite index $tab$, the linear system solved by the $\alpha$ is simply
\begin{equation}
    C = \mathcal{V} \alpha ~ .
\end{equation}
Now, identical arguments apply as in the scalar case: the $\alpha_k$ are uniquely defined so long as $\mathcal{V}$ is of full column rank (i.e., has rank $rm$).
The condition $\text{rank}(\mathcal{V}) = rm$ is therefore the additional assumption required beyond invertibility of $\bm{H}^{(0)}$ to guarantee uniqueness of the $\bm{a}_k$.
Numerical checks with lattice and random data suggest that violations of this condition are improbable with noisy data.

\subsection{Companion matrices}
\label{sec:companion}

We can obtain a different but equivalent view of Prony's method by defining the $m \times m$ \emph{companion matrix} associated with the (monic) polynomial $p^{(m)}(\lambda)$,
\begin{equation}
\label{eq:companion-scalar-def}
    \mathcal{C} \equiv \begin{bmatrix}
        0 & 0 & \cdots & 0 & \gamma_0 \\
        1 & 0 & \cdots & 0 & \gamma_1 \\
        0 & 1 & \cdots & 0 & \gamma_2 \\
        \vdots & \vdots & \ddots & \vdots & \vdots \\
        0 & 0 & \cdots & 1 & \gamma_{m-1}
    \end{bmatrix} ~ .
\end{equation}
By construction, $\det[\lambda \mathbb{I} - \mathcal{C}] = p^{(m)}(\lambda)$, where $\mathbb{I}$ is an $m\times m$ identity matrix, so the eigenvalues of $\mathcal{C}$ coincide with the roots of $p^{(m)}$, i.e., with $\lambda_k$.
It follows from \cref{eq:prony-scalar-recurrence} that the action of $\mathcal{C}^T$ on vectors of correlator data $\vec{C}^{(t_0)} \equiv [C(t_0), \ldots, C(t_0+m-1)]$ is to increment $t_0$, i.e., 
\begin{equation}\label{eq:scalar-companion-on-Cvec}
    \mathcal{C}^T \vec{C}^{(t_0)} = \vec{C}^{(t_0+1)}
\end{equation}
for any $t_0 \in \{0,\ldots,m-1\}$.
Its action on Hankel matrices is similar, i.e.,
\begin{equation}\label{eq:scalar-companion-hankel-action}
    H^{(t_0)} \mathcal{C} = H^{(t_0 + 1)} 
    ~~\text{ and }~~
    \mathcal{C}^T H^{(t_0)} = H^{(t_0 + 1)} ~ .
\end{equation}
We thus see that, if $H^{(0)}$ is invertible, we can compute the companion matrix as $\mathcal{C} = [H^{(0)}]^{-1} H^{(1)}$, where $H^{(0/1)}$ are both $m \times m$.
This provides an alternate view on uniqueness of the $\lambda_k$, which here follows simply from uniqueness of the eigendecomposition of $\mathcal{C}$.

The companion matrix admits straightforward generalization to the block case.
Just as the $\gamma_s$ admit two distinct generalizations into the matrices $\bm{\Gamma}^{R/L}_\sigma$, there are distinct $L$ and $R$ block generalizations to consider:
\begin{equation}
\label{eq:companion-block-right-def}
    \bm{\mathcal{C}}^R = \begin{bmatrix}
        0 & 0 & \cdots & 0 & \bm{\Gamma}^R_0 \\
        \bm{1} & 0 & \cdots & 0 & \bm{\Gamma}^R_1 \\
        0 & \bm{1} & \cdots & 0 & \bm{\Gamma}^R_2 \\
        \vdots & \vdots & \ddots & \vdots & \vdots \\
        0 & 0 & \cdots & \bm{1}  & \bm{\Gamma}^R_{m-1}
    \end{bmatrix} ~ ,
\end{equation}
and
\begin{equation}
\label{eq:companion-block-left-def}
    \bm{\mathcal{C}}^L = \begin{bmatrix}
        0 & \bm{1} & 0 & \cdots & 0 \\
        0 & 0 & \bm{1} & \cdots & 0 \\
        \vdots & \vdots & \vdots & \ddots & \vdots \\
        0 & 0 & 0 & \cdots & \bm {1} \\
        \bm{\Gamma}^L_0 & \bm{\Gamma}^L_1 & \bm{\Gamma}^L_2 & \cdots & \bm{\Gamma}^L_{m-1}
    \end{bmatrix} ~ .
\end{equation}
By construction, $\det[\lambda \bm{1} - \bm{\mathcal{C}}^{R/L}] = \det \bm{p}^{R/L}(\lambda)$, so the eigenvalues of $\bm{\mathcal{C}}^{R/L}$ are the same as the roots of $\det \bm{p}^{R/L}(\lambda)$.
Similarly as in the scalar case, the block companion matrices $\bm{\mathcal{C}}^{R/L}$ relate the block Hankel matrices $\bm{H}^{(0)}$ and $\bm{H}^{(1)}$ as
\begin{equation}\label{eq:block-companion-hankel-action}
    \bm{H}^{(0)} \bm{\mathcal{C}}^R = \bm{H}^{(1)}
    ~~\text{ and }~~
    \bm{\mathcal{C}}^L \bm{H}^{(0)}  = \bm{H}^{(1)} ~ .
\end{equation}
Assuming $\bm{H}^{(0)}$ is invertible, it follows from \cref{eq:block-companion-hankel-action} that 
\begin{equation}
    \bm{\mathcal{C}}^R = [ \bm{H}^{(0)}]^{-1} \bm{\mathcal{C}}^L \bm{H}^{(0)} ~ ,
\end{equation}
i.e., that $\bm{\mathcal{C}}^R$ and $\bm{\mathcal{C}}^L$ are similar.
This means that their eigenvalues---and thus the roots of $\det \bm{p}^R(\lambda)$ and $\det \bm{p}^L(\lambda)$---coincide.

\section{The Prony-Ritz Equivalence Class}
\label{sec:prony-ritz}

This section characterizes how various different approaches to correlator analysis fit within a broader equivalence class, which we call the ``Prony-Ritz equivalence class''.
The defining criterion for methods in this class is that they produce exact correlator decompositions.
When uniqueness holds, all methods in the class will necessarily produce identical unfiltered outputs when applied to generic correlator data (more precisely, any correlator data for which the Prony roots are non-degenerate, and the Hankel matrix and generalized Vandermonde matrix are non-singular). 

The section provides additional details on how correlator analysis methods relate to the Prony-Ritz equivalence class and further discussion of the implications of uniqueness.
First, we note that previous identifications between various correlator analysis methods in Refs.~\cite{Fischer:2020bgv,Wagman:2024rid,Ostmeyer:2024qgu,Chakraborty:2024scw} can be understood simply from the appearance of the Prony companion matrix in each method.
Second, we note the limitations of equivalence---specifically, that it holds only at the level of unprocessed outputs before the effects of noise are treated.
In this light, we can understand what is novel about a FRR analysis in comparison to early application of Prony's method~\cite{Fleming:2004hs,Lin:2007iq,Fleming:2009wb,Beane:2009kya} and GPOF~\cite{Aubin:2010jc,Aubin:2011zz,Green:2014xba,Schiel:2015kwa,Ottnad:2017mzd,Fischer:2020bgv} to LQCD: a principled approach to treatment of noise artifacts that retains optimality guarantees and allows convergence properties to be understood.
Third, a FRR analysis also admits a vector-space picture which allows better understanding how the data relate to and constrain the properties of physical states.
Finally, we close with a taxonomy of different analysis methods and how they relate to RR.

\subsection{The companion matrix \& direct equivalences}
\label{sec:companion-equiv}

The central object to more direct coincidences between Prony, RR, and GPOF  is the companion matrix and its block generalizations, as introduced in \cref{sec:companion}.
In block Prony, the $R/L$ companion matrices $\bm{\mathcal{C}}^{R/L}$ can be computed from \cref{eq:block-companion-hankel-action} as $\bm{\mathcal{C}}^R = [\bm{H}^{(0)}]^{-1} \bm{H}^{(1)}$ and $\bm{\mathcal{C}}^L = \bm{H}^{(1)} [\bm{H}^{(0)}]^{-1}$.
Note that in block Prony with $r=1$, both the $R/L$ companion matrices have the same spectrum as the usual scalar companion matrix, $\mathcal{C}^R = [\mathcal{C}^L]^T = \mathcal{C}$, and so it is sufficient to discuss block and treat scalar as the special case $r=1$. 

The block companion matrices $\bm{\mathcal{C}}^R$ and $\bm{\mathcal{C}}^L$ correspond exactly to matrix representations of $T^{(rm)}$ as computed by ORR with different oblique conventions.
Comparison of $\bm{\mathcal{C}}^R = [\bm{H}^{(0)}]^{-1} \bm{H}^{(1)}$ with \cref{eq:Hankel_Tm} in the form
\begin{equation}
    \bm{T}^{(rm)} = [\bm{E}^L]^{-1} \bm{H}^{(1)} [\bm{E}^R]^{-1}
\end{equation}
shows that $\bm{\mathcal{C}}^R$ is equal to $\bm{T}^{(rm)}$ with oblique conventions such that $\bm{E}^L = \bm{H}^{(0)}$ and $\bm{E}^R = \bm{1}$.
Similarly, $\bm{\mathcal{C}}^L$ is equal to $\bm{T}^{(rm)}$ with oblique conventions such that $\bm{E}^R = \bm{H}^{(0)}$ and $\bm{E}^L = \bm{I}$.
We thus see that Prony roots (the eigenvalues of $\bm{\mathcal{C}}^R$ or $\bm{\mathcal{C}}^L$) are Ritz values (the eigenvalues of $T^{(rm)}$).
When (block) Prony is implemented via companion matrices and $\bm{H}^{(0)}$ is invertible, it is an identical algorithm to ORR before filtering.
This is an even stronger statement than equivalence between (oblique, block) Lanczos and (O,B)KRR---in the usual definition of Lanczos, $\bm{T}^{(rm)}_{ij}$ is evaluated in a different basis, and although it has the same spectrum its matrix elements are not identical to those of $\bm{\mathcal{C}}^R_{ij}$ or $\bm{\mathcal{C}}^L_{ij}$.

GPOF prescribes obtaining the eigenvalues by solving either the right or left GEVP,
\begin{equation}
    \bm{H}^{(1)} \vec{g}^R_k = \bm{H}^{(0)} \vec{g}^R_k \lambda_k
    ~~\text{ or }~~
    \vec{g}^{L\dagger}_k \bm{H}^{(1)} = \lambda_k \vec{g}^{L\dagger}_k \bm{H}^{(0)} ~ .
\end{equation}
When $\bm{H}^{(0)}$ is invertible, a valid algorithm to solve these GEVPs is simply to invert $\bm{H}^{(0)}$ to obtain the equivalent standard eigenproblems and solve those instead.
Identifying the block companion matrices, they are
\begin{equation}
    \bm{\mathcal{C}}^R \vec{g}^R_k = \vec{g}^R_k \lambda_k
    ~~\text{ and }~~
    \vec{g}^{L\dagger}_k \bm{\mathcal{C}}^L = \lambda_k \vec{g}^{L\dagger}_k ~ .
\end{equation}
Implemented in this manner, GPOF reduces to computing the eigenvalues of $\bm{\mathcal{C}}^{R/L}$ and is similarly identical to ORR before filtering.
In this case, the coincidence goes beyond just eigenvalues: as shown in \cref{sec:gevp}, the generalized eigenvectors used to compute GPOF overlaps and matrix elements~\cite{Aubin:2010jc,Aubin:2011zz,Green:2014xba,Ottnad:2020qbw,Ottnad:2017mzd} are exactly the Ritz coefficients $P^{R/L(m)}$ from RR.
This applies also to the GEVP versions thereof obtained in the two-timeslice limit of GPOF~\cite{Blossier:2009kd,Dragos:2016rtx,Bulava:2016mks}; the coincidence between one step of block Lanczos(=RR) and GEVP was already noted in Ref.~\cite{Hackett:2024nbe}.

\subsection{Equivalence and its limitations}

It is worth emphasizing that equivalences between Prony, Lanczos, GPOF, and other methods in the Prony-Ritz equivalence class hold at the level of unfiltered outputs, before applying any postprocessing or statistical interpretations.
In \cref{sec:prony-uniqueness}, we explicitly construct an invertible map between the $\lambda_k$ and $\bm{a}_k$ and the correlator data $\bm{C}(t)$ that is analogous to a change-of-basis.\footnote{For each $k$, there are $2r-1$ degrees of freedom in $\bm{a}_k = \vec{z}_k^L \vec{z}_k^{R\dagger}$ due to the overall normalization ambiguity discussed above. The product $\lambda_k \bm{a}_k$ therefore has $2r$ degrees of freedom. Since there are $mr$ states, $\{ \lambda_k, \bm{a}_k \} $ has the $2mr^2$ degrees of freedom required to be in one-to-one correspondence with the $2mr^2$ correlator data, $\{ \bm{C}(t) \}$. See Ref.~\cite{Fleming:2023zml} for further discussion. }

From this perspective, it is clear that determination of the  $\{ \lambda_k, \bm{a}_k \} $ from the $\{  \bm{C}(t)  \} $ is a unique transformation that can be performed using whatever means is convenient.
Any difference between different correlator analysis approaches in the Prony-Ritz equivalence class necessarily arises in postprocessing and interpretation of the $\{ \lambda_k, \bm{a}_k \}$---i.e., how these quantities are refined into estimators for energies and overlaps.
Before this stage, any method in the equivalence class---including (block) Lanczos, (block) Prony, and GPOF---produces equivalent outputs.\footnote{Other analysis methods exist which lie outside the equivalence class. Some methods which violate the strict definition considered here may nevertheless be equivalent for all practical purposes; for example, multi-state fits with enough degrees of freedom to describe noisy data almost exactly. }

Nontrivial numerical differences may arise between the different methods in postprocessing.
In particular, each method is traditionally accompanied by different standard prescriptions for removing or otherwise regulating the effects of the spurious eigenvalues (spurious states, in Lanczos language) that always arise when analyzing sufficiently long time series of noisy correlator data.
These different regulation prescriptions lead to different outputs and so, as applied in practice, these methods have different performance characteristics.
As discussed and emphasized in \cref{sec:filtering}, the Lanczos viewpoint provides a particularly simple and well-motivated approach to removing noise artifacts through spurious-state filtering.
Due to uniqueness of the $\{ \lambda_k, \bm{a}_k \} $, Lanczos technology for spurious-state filtering and residual-bound calculation may be combined straightforwardly with any other methods in the Prony-Ritz equivalence class through identification between the companion matrices and $T^{(rm)}$ discussed above and/or the relations discussed in \cref{sec:rr-practice}.

\subsection{Vector-space interpretations}

Further distinctions between methods arise in the interpretation of the data.
The correlator decomposition and Prony's method are motivated purely algebraically, and thus may be applied to any data which can be expected to have an exponential decomposition.
However, as noted in \cref{sec:rr-practice}, the
ORR method naturally introduces an $rm$-dimensional vector space where $T^{(rm)}$ is represented by an explicit matrix.
The vector-space description is not unique, and holds for any choices of $T^{(rm)}$, $\ket{\psi_a}$, and $\bra{\chi_a}$ which numerically reproduce the correct $\bm{a}_k$ and $\lambda_k$.
For example, there is freedom to perform both change-of-bases as well as joint rescaling of the $\ket{\psi_a}$ and $\bra{\chi_a}$.
Thus, caution is required when associating these quantities with (approximations of) the properties of physical states.

We can explore the space of possible interpretations by considering another framing of Prony methods:
they proceed by characterizing the correlator data as a solution to an $rm$th-order recurrence relation or ``linear difference equation'' (LDE), the discrete analog of an ordinary differential equation.
It is a well-known fact that any solution to such an LDE admits description non-uniquely as
\begin{equation}\label{eq:prony-recurrence-description}
    C(t) = u^{L\dagger} A^t u^R
\end{equation}
in terms of a square matrix $A$ and pair of column vectors $u^R$ and $u^L$ with dimension at least $rm$.
The structural connection with Lanczos methods is immediately apparent.
Here, we explore it further to highlight the distinction between purely algebraic analyses and the vector-space interpretation offered by Rayleigh-Ritz methods, as well as important subtleties in this interpretation.

As already noted, the description \Cref{eq:prony-recurrence-description} is not unique.
Restricting to $rm$ dimensions, we find various options all related by changes of basis.
For example, if $A$ is taken as $\bm{\mathcal{C}}^{R}$, then
\begin{equation}
    \bm{C}(t) =
    \begin{bmatrix}
        \bm{C}(0) & \cdots & \bm{C}(m-1)
    \end{bmatrix}
    ~ \bm{\mathcal{C}^R} ~ 
    \begin{bmatrix}
        \bm{1} \\ \bm{0} \\ \vdots \\ \bm{0}
    \end{bmatrix} ~ .
\end{equation}
We can write an equivalent description in the eigenbasis of $\bm{\mathcal{C}}^{R}$, or equivalently $T^{(rm)}$ in any other oblique convention, simply by inserting $T^{(rm)} = \omega^{(rm)} \Lambda^{(rm)} (\omega^{-1})^{(rm)}$ and recognizing definitions:
\begin{equation}
\begin{aligned}
    C_{ab}(t) = &
    \begin{bmatrix}
        Z^{R(rm)*}_{0,a} & \cdots & Z^{R(rm)*}_{rm-1,a}
    \end{bmatrix} \times
    \\ & \qquad 
    \begin{bmatrix}
        \lambda^{(rm)}_0  &  & 0 \\
         & \ddots &  \\
        0 &  & \lambda^{(rm)}_{rm-1}
    \end{bmatrix}^t
    \begin{bmatrix}
        Z^{L(rm)}_{0,b} \\ \vdots \\ Z^{L(rm)}_{rm-1,b}
    \end{bmatrix} ~ .
\end{aligned}
\end{equation}
The result is simply the correlator decomposition written in matrix-vector form.
Uniqueness of the decomposition implies that any other choice of $rm$-dimensional description is equivalent up to a change of basis, which corresponds to the freedom to choose an oblique convention in ORR.

However, while \Cref{eq:prony-recurrence-description} requires at least $rm$ dimensions to hold, we are free to consider descriptions in more dimensions.
Taking this to its logical extreme, we may write a description $C_{ab}(t) = \braket{u^L_a | A^t | u^R_b}$ in terms of an operator $A$ and states $\bra{u^L_a}$ and $\ket{u^R_b}$ in an infinite-dimensional Hilbert space.
Because a minimal description of $\bm{C}(t)$ requires only an $rm$-dimensional vector space, any Hilbert-space description is far from unique.
Different descriptions are no longer related by simple basis changes.
Formally speaking, further input is required to relate the data to the Hilbert space of physical states.

A theoretical advantage of the Rayleigh-Ritz formalism presented herein over purely algebraic methods is a natural characterization of the data in terms of states and operators acting in subspaces of Hilbert space.
A Hermitian subspace can be defined in which Ritz vectors may be interpreted as approximations to physical energy eigenvectors.
The remainder of the Ritz vectors can be characterized as noise artifacts.
This physical characterization of the information that can be extracted from lattice correlators is useful for defining matrix-element estimators~\cite{Hackett:2024xnx} and motivating spurious-state filtering methods such as the ZCW test~\cite{Hackett:2024nbe}; it may also prove useful in a variety of other future applications.

However, it is important to note that caution is required in this interpretation.
Because it is an interpretation imposed by the analyzer and not by the numerical data, it is possible to obtain misleading results from naive application.
For example, consider an off-diagonal correlator $C(t) = \braket{\psi^L|T^t|\psi^R}$ for which all amplitudes $a_k = \braket{\psi^L|y^{R(m)}_k}^* \braket{y^{L(m)}_k|\psi^R}$ are positive.
Such a correlator admits a numerically equivalent ``fictitious diagonal'' description as $C(t) = \braket{\tilde{\psi} | \tilde{T}^t | \tilde{\psi}}$.
If analyzed exactly as a diagonal correlator, no obvious internal inconsistencies will arise in the numerics, but the values of quantities like $\braket{\tilde{\psi} | \tilde{y}^{R(m)}_k}$ and $\braket{\tilde{\psi} | \tilde{y}^{L(m)}_k}$ will be unrelated to the physically meaningful overlaps $\braket{\psi^L|y^{R(m)}_k}$ and $\braket{y^{L(m)}_k|\psi^R}$.
Carefully distinguishing right and left quantities is critical to a proper interpretation.

Separately, while RR methods may be applied to any data admitting an exponential decomposition, the motivation underlying machinery like Hermitian subspace and ZCW filtering applies only for correlator data where identification with the physical space of states is possible.
Other obstacles arise for instance in applications of RR to calculations of matrix elements employing the summation method, Feynman-Hellman method, and related techniques as discussed in \cref{app:feynhell}.

\subsection{Taxonomy}
\label{subsec:tax}

\begin{table*}
\begin{ruledtabular}
\begin{tabular}{cccrc}
    \textbf{Hermiticity} & \textbf{Block} & \textbf{Convergence} & \textbf{RR Name} &  \textbf{Correlator analysis methods} \\\hline
    Yes & No & Power iteration & RR & Effective masses, ratios, etc. \\
    Yes & No & KPS & KRR & Lanczos, Prony, GPOF/PGEVM/TGEVP \vspace{0.8em}\\
    Yes & Yes & Power iteration & RR & GEVP methods \\
    Yes & Yes & KPS & BKRR & Block Lanczos, Block Prony, GPOF \vspace{0.8em}\\
    No & No & Power iteration & ORR & Effective masses, ratios, etc. \\
    No & No & Oblique & OKORR & Oblique Lanczos, Prony, GPOF/PGEVM/TGEVP \\
    No & No & KPS$^*$ & FOKORR & Filtered [Oblique Lanczos, Prony, GPOF/PGEVM/TGEVP] \vspace{0.8em}\\
    No & Yes & Power iteration & ORR & GEVP methods \\
    No & Yes & Oblique & OBKORR & Oblique Block Lanczos, Block Prony, GPOF \\
    No & Yes & KPS$^*$ & FOBKORR & Filtered [Oblique Block Lanczos, Block Prony, GPOF]
\end{tabular}
\end{ruledtabular}
  \caption{Taxonomy of various methods for correlator analysis as they relate to the Rayleigh-Ritz method; see \cref{subsec:tax}. The first two columns refer to whether the transfer matrix that exactly describes the (noisy) correlator as a time series is Hermitian and whether the correlator is matrix-valued, respectively. The third column describes the convergence of the RR method to be applied;  KPS$^*$ convergence refers to KPS convergence within the Hermitian subspace as described in \cref{sec:filtering-convergence}. 
\label{tab:tax}
  }
\end{table*}

As shown in \cref{tab:tax}, a wide range of methods can be understood as implicitly applying RR to different subspaces.
Key properties of the methods follow from the type of the subspace as indicated in \cref{tab:tax}.
Some methods assume Hermiticity of an underlying transfer matrix representation of correlator data (Hermiticity = yes), while others do not.
Similarly, some methods are applicable to correlator matrices (block = yes), while others are explicitly defined for the scalar case.
For data that can be exactly described by Hermitian $T^{(rm)}$,  Krylov  subspace methods achieve faster convergence (KPS) than iterative applications of RR to BK subspaces of the same dimension (power iteration).
For data described  by non-Hermitian $T^{(rm)}$, Krylov subspace methods have different convergence properties (oblique).
Other correlator analysis methods have been proposed that are closely related to RR methods but are not equivalent to application of RR to a subspace of Hilbert space and therefore do not fall precisely within this taxonomy:

\begin{itemize}[leftmargin=*]
\item The matrix-Prony method proposed in Ref.~\cite{Beane:2009kya} first introduced the connection between Prony's method and the GEVP appearing in GPOF/PGEVM/TGEVP.
Matrix-Prony is identical to GPOF for time series with length two.
In general, matrix-Prony provides a version of GPOF where only certain rows of the Hankel matrices are computed and the solutions can be described as least-squares approximations to the underdetermined GEVP involving these rectangular matrices.
Matrix-Prony is explicitly defined for vectors of scalar correlators and can therefore be thought of as a generalization of OKORR in the style of GPOF that is applicable in situations where not all the data required to construct square Hankel matrices is available.

\item The ``PGEVM/GEVM'' method proposed in Ref.~\cite{Fischer:2020bgv} analyzes rank-$r$ correlator matrices via two sequential applications of RR in dimension $r$ and dimension $m$ spaces.
The first application uses a subspace of fixed dimension, while the second application involves Krylov spaces of increasing dimension.
This corresponds to $r$ distinct applications of dimension-$m$ OKORR, which leads to slower convergence than a single dimension-$rm$ application of OBKORR as discussed in Ref.~\cite{Hackett:2024nbe}. 
\end{itemize}

\section{Conclusions}
\label{sec:conclusions}

We have investigated coincidences between methods previously applied to LQCD correlator analysis including (block) Prony's method, (block) Lanczos, GPOF, and GEVP methods.
Some of these coincidences have been previously proven or conjectured.
Here, we have proven the general result that any method for obtaining an exact representation of (matrix) correlators as sums of real or complex exponentials provides identical transfer-matrix eigenvalue and eigenvector approximations: the Ritz values and vectors.
This criterion, which defines the Prony-Ritz equivalence class, provides a simple practical test for whether other methods are equivalent.

For real, symmetric correlator matrices, Ritz values are optimal approximations to transfer-matrix eigenvalues within a Krylov space naturally associated with applications of the transfer matrix to an initial-state subspace.
Krylov-space methods converge exponentially faster than power-iteration methods, which include effective masses as well as GEVP methods for correlator matrices, as quantified by Kaniel-Paige-Saad convergence theory.
All of the correlator analysis methods discussed here are equivalent to  applications of the Rayleigh-Ritz method to Krylov spaces or other subspaces when applied to data with a manifestly Hermitian spectral representation.

When applied to noisy LQCD correlator data or other time series without a manifestly Hermitian spectral representation, correlator analysis methods cannot be interpreted strictly as applying RR to block Krylov spaces.
They must be understood as applying an oblique version of RR, such as oblique Lanczos, with different convergence properties.
However, filtering methods such as the Cullum-Willoughby test, the $R/L$ norm trick, and the ZCW test can be applied to noisy approximations of real, symmetric correlator matrices.
By restricting to a Hermitian subspace spanned by Ritz vectors that can be interpreted as physical states, it is possible to view filtered RR as application of (non-oblique) RR to the space of non-spurious states. 
This means that FRR provides optimal approximations to transfer matrix eigenvalues within a physically interpretable subspace, even in the presence of noise.

We further proposed numerical implementations of OBKORR based on directly solving the Rayleigh-Ritz problem in Krylov space and by solving the equivalent GEVP involving block Hankel matrices at pairs of adjacent timeslices.
In the scalar case, the Hankel GEVP version of OKRR corresponds to solving precisely the same GEVP as in GPOF/PGEVM/TGEVP, although the prescribed usage here is to include increasingly many correlator timeslices rather than a fixed-length sliding window.
The block Hankel GEVP version of OBKORR involves the same GEVP as in the general case of GPOF.

Advantages of different perspectives can be combined by mixing tools for calculating and interpreting Ritz values and vectors that were developed in different contexts.
The advantages provided by the OBKORR (i.e. oblique block Lanczos) perspective in comparison with Prony's method and GPOF are theoretical tools available from Krylov space including KPS convergence theory, directly computable residual bounds, and the Hilbert-space picture underlying spurious-state filtering methods such as the ZCW test. 
In comparison with oblique block Lanczos, both the direct ORR and block Hankel GEVP (i.e.~GPOF) implementations of OBKORR are simpler to implement and more 
numerically stable.
We find this allows working entirely in double precision rather than the higher precision required for Lanczos recursions, allowing faster and more parallelizable implementation in practice.
Further detailed comparisons of the numerical stability of different implementations will be an important topic for future work.

Future applications can achieve both sets of advantages by combining numerically stable and efficient GPOF-style OBKORR implementations with spurious-state filtering and convergence/error analysis based on an oblique block Lanczos interpretation of the results. 
This will be particularly advantageous for 
applications of FRR to very large correlator matrices, such as those arising in lattice QCD calculations of hadron-hadron scattering and electroweak transitions involving resonances.

\begin{acknowledgements}
This document was prepared using the resources of the Fermi National Accelerator Laboratory (Fermilab), a U.S. Department of Energy, Office of Science, Office of High Energy Physics HEP User Facility. Fermilab is managed by Fermi Forward Discovery Group, LLC, acting under Contract No. 89243024CSC000002.
RA is supported in part by the U.S.\ Department of Energy, Office of Science, Office of Nuclear Physics, under grant Contract Number DE-SC0011090, and also by the High Energy Physics Computing Traineeship for Lattice Gauge Theory (DE-SC0024053). DAP is supported by the U.S. Department of Energy (DOE), Office of Science, Office of Nuclear Physics, under grant contract number DE-SC0004658.
\end{acknowledgements}

\appendix

\section{Notation \& Glossary}
\label{sec:glossary}

This appendix summarizes notation and conventions used throughout the main text.

Boldfaced symbols like $\bm{C}(t)$ are shorthand for matrices $C_{ab}(t)$ in the  missing indices $a,b$. Boldface is used for both block indices and generic RR indices, the difference being clear from context.
Matrix components are products are denoted by
\begin{equation}
    [\bm{a}_j]_{ab} \equiv \alpha_{jab}
    \quad \text{or} \quad 
    \left[ \bm{A} \bm{B} \right]_{ab} \equiv \sum_c A_{ac} B_{cb} ~ .
\end{equation}
Bolded symbols are always matrices and never vectors.
(Column) vectors are denoted with vector hats, e.g., $g_{ik} = [\vec{g}_k]_i$.
Where it benefits clarity, this convention is in some cases applied inconsistently; in particular, in \cref{sec:prony-uniqueness}, bolding is reserved to distinguish block quantities from scalar ones.

The dimensions of various spaces are denoted by:
\begin{itemize}
    \item $d$ denotes the dimension of a generic subspace.
    \item $m$ denotes the dimension of Krylov space and/or the number of steps in applications of iterative methods.
    \item $r$ denotes the block dimension, i.e.~the rank of a correlator matrix.
\end{itemize}
To match conventions elsewhere in the literature, unavoidably some indices are conventionally zero-indexed and others are conventionally one-indexed:
\begin{itemize}
    \item $a,b,c,d \in \{0,\ldots,r-1\}$ are reserved for block indices corresponding to the indices on correlator matrices $C_{ab}(t)$.
    \item $k,l \in \{0,\ldots,d-1\}$ are reserved for the state index on Ritz values $\lambda^{(d)}_k$ and vectors $\ket{y^{R/L(d)}_k}$.
    \item $n \in \{0, \ldots \}$ is reserved for the state index on true eigenvalues and eigenvectors.
    \item $i,j$ are one-indexed. They are employed for the generic ORR basis vectors $\ket{\xi^{R/L}_i}$ with $i \in \{1,\ldots,d\}$, as well as intermediate basis vectors when they have block indices (of which the block Lanczos vectors are a special case), $\ket{v^{R/L}_{ia}}$, with $i \in \{1,\ldots,m\}$.
    \item $t,s,\tau,\sigma \in \{0,\ldots,2m-1\}$ are reserved for Euclidean times, as in block Krylov vectors $\ket{k^{R/L}_{ta}}$.
\end{itemize}
Correlators and correlator matrices are denoted
\begin{itemize}   
    \item $C(t) = \braket{\psi^L|T^t|\psi^R} = \sum_k a_k \lambda_k^t$ is a correlator, which is defined for integer values of its Euclidean-time argument $t$.
    \item $[\bm{C}(t)]_{ab} = \braket{\psi^L_a | T^t | \psi^R_b} = \sum_k a_{k,ab} \lambda_k^t$ is a matrix of correlation functions.
    \item $H^{(p)}_{st} = C(s+t+p)$ is a Hankel matrix.
    \item $\bm{H}^{(p)}_{st} = \bm{C}(s+t+p)$ is a block Hankel matrix.
\end{itemize}
The transfer matrix, its spectrum, and ORR approximations to it involve:
\begin{itemize}
    \item $T$ is the transfer matrix of the lattice field theory.
    \item $\lambda_k$ denotes the $k$th eigenvalue of the transfer matrix $T$, the decay constant of the $k$th term appearing in a correlator decomposition $\sim \lambda_k^t$, or the $k$th Prony root.
    \item $\ket{n}$ and $\bra{n}$ denote the $n$th true eigenstate of the full transfer matrix $T$.
    \item $T^{(d)}$ denotes the transfer matrix projected to a $d$-dimensional subspace.
    \item $\lambda^{(d)}_k$ denotes the $k$th Ritz value computed in a $d$-dimensional subspace.
    \item $\ket{y^{R(d)}_k}$ and $\bra{y^{L(d)}_k}$ denote the $k$th Ritz vectors computed in a $d$-dimensional subspace.
\end{itemize}
Vector spaces arising in ORR are:
\begin{itemize}
\item $\mathcal{H}$ is the Hilbert space of states.
    \item $\mathcal{S}^{(d)}$ is a $d$-dimensional subspace of $\mathcal{H}$.
    \item $\mathcal{K}^{(m)}(\ket{\psi}) = \Span[\ket{\psi}, T \ket{\psi}, \ldots, T^{m-1} \ket{\psi}]$ is the Krylov space for vector $\ket{\psi}$ and matrix $T$.
    \item $\mathcal{K}^{(rm)}(\mathcal{S}^{(r)})$ is the block Krylov space for initial vectors spanning a subspace $\mathcal{S}^{(r)}$.
    \item $\herm$ is the Hermitian subspace defined by the span of $T^{(d)}$ eigenvectors for which $T^{(d)}$ acts as a Hermitian operator.
    \item $\overline{\mathfrak{s}}$ is the non-spurious subspace defined by a particular filtering prescription.
\end{itemize}
Vectors arising in ORR are:
\begin{itemize}
    \item Kets $\ket{\psi^R_a}$ and bras $\bra{\psi^L_a}$ denote the ``initial/final states'' in the (infinite-dimensional) Hilbert space of states, excited by the corresponding interpolating operators $\psi^{R\dagger}$ and $\psi^L$.
    \item $\ket{\xi^R_i}$ and $\bra{\xi^L_j}$ are generic basis vectors in ORR.
    \item $\ket{k^R_{ta}} = T^t \ket{\psi^R_a}$ and $\bra{k^L_{ta}} = \bra{\psi^L_a} T^t$ are Krylov vectors.
    \item $\ket{v^R_i}$ and $\bra{v^L_j}$ are biorthonormal bases satisfying $\braket{v^L_i|v^R_j} = \delta_{ij}$ constructed such that $\Span\aset{\ket{\xi^R_{i}}} = \Span\aset{\ket{v^R_i}}$ and $\Span\aset{\bra{\xi^L_{i}}} = \Span\aset{\bra{v^L_i}}$.
\end{itemize}
Important matrices involving these vectors are:
\begin{itemize} 
    \item $\bm{M}^{(rm)} = \bm{H}^{(1)}$ is the matrix representation of $T$ in the Krylov basis,  $[\bm{M}]_{sa,tb} = \braket{k^L_{sa} | T | k^R_{tb}} $.
    \item $\bm{G}^{(rm)} = \bm{H}^{(0)}$ is the ``metric'' of inner products of Krylov vectors, $[\bm{G}]_{sa,tb} = \braket{k^L_{sa} | k^R_{tb}} $.
    \item $\bm{E}^R$  and $\bm{E}^L$ satisfying $G^{(rm)}_{sa,tb} = \sum_{i} E^L_{sa,i} E^R_{i,tb}$ are factors of the metric whose specification defines an oblique convention.
    \item  $K^R_{ta,i} = [(\bm{E}^R)^{-1}]_{ta,i}$ and $K^L_{i,ta} = [(\bm{E}^L)^{-1}]_{i,ta}$ are ``Krylov coefficients'' providing change-of-basis matrices connecting the $\ket{k^{R/L}_{ta}}$ and $\ket{v_i^{R/L}}$.
    \item $\omega^{(rm)}_{jk}$ are eigenvectors of $T^{(rm)}$ providing change-of-basis matrices for the $\ket{v_i^{R/L}}$ and $\ket{y_k^{R/L(rm)}}$.
    \item  $\mathcal{N}_k^{(rm)}$ are normalization coefficients defined so that $\braket{y_k^{(rm)R/L} | y_k^{(rm)R/L}} = 1$.
    \item $P^{R(rm)}_{ta,k} = \mathcal{N}_k^{(rm)} \sum_j K^R_{ta,j} \omega^{(rm)}_{jk} $ and\\ $P^{L(rm)}_{k,ta} = 1/ \mathcal{N}_k^{(rm)} \sum_j  (\omega^{-1})^{(rm)}_{kj} K^L_{j,ta} $ are Ritz coefficients, the change-of-basis matrices connecting the $\ket{k^{R/L}_{ta}}$ and $\ket{y_k^{R/L(rm)}}$.
\end{itemize}
Amplitudes appearing in correlator decompositions are:
\begin{itemize} 
    \item $Z_{ka} \equiv \braket{n | \psi_a}$ is an overlap factor.
    \item $a_k = Z^{R*}_k Z^L_k$ is an amplitude as appears in a scalar correlator decomposition.
    \item $[\bm{a}_k]_{ab} = a_{k,ab} = Z^{R*}_{ka} Z^L_{kb}$ is a rank-one amplitude matrix as appears in a correlator matrix decomposition.
    \item $\tilde{\bm{a}}_k \propto \bm{a}_k$ is an unnormalized amplitude matrix, related to $\bm{a}_k$ by an overall complex constant.
    \item $Z^{R(d)}_{ka} = \braket{y^{R(d)}_k | \psi_a}$ and $Z^{L(d)}_{ka} = \braket{y^{L(d)}_k | \psi_a}$ denote overlap factors for a Ritz vector, rather than a true eigenstate.
\end{itemize}
Additional algebraic structures that arise in Prony's method and the Prony-Ritz uniqueness proof are:
\begin{itemize} 
    \item $p^{(d)}(\lambda) = \prod_k (\lambda - \lambda_k)$ is a $d$th-order characteristic polynomial.
    \item $\gamma_s$ are the linear predictions coefficients, which are the coefficients of the characteristic polynomial as in $p^{(d)}(\lambda) = \lambda^d - \sum_s \gamma_s \lambda^s$.
    \item $\bm{p}^{R(d)}$ and $\bm{p}^{L(d)}$ are matrix generalizations of the characteristic polynomial.
    \item $\bm{\Gamma}^R_\sigma$ and $\bm{\Gamma}^L_\sigma$ are the linear prediction matrices which generalize $\gamma_s$ to the block case.
    \item $V_{tk} = \lambda_k^t$ is the Vandermonde matrix. Note this definition is transposed from the standard one.
    \item $\mathcal{V}_{tk} = V_{tk} \tilde{\bm{a}}_k$ is a generalization of the Vandermonde matrix.
    \item $\mathcal{C}$ is the companion matrix.
    \item $\bm{\mathcal{C}}^R$ and $\bm{\mathcal{C}}^L$ are block companion matrices.
\end{itemize}

\begin{widetext}

\section{Residual bound derivation}
\label{app:res-bounds}

The Lanczos residual bound~\cite{Parlett} applies to oblique Lanczos when $T = T^\dagger$ even when $T^{(rm)} \neq [T^{(rm)}]^\dagger$~\cite{Wagman:2024rid}. 
The corresponding residual bound for oblique block Lanczos takes the form~\cite{Hackett:2024nbe}
\begin{equation}
    \mathrm{min}_{\lambda \in \aset{\lambda_n}} |\lambda - \lambda^{(rm)}_k|^2
    \leq |B^{R/L(rm)}_k|,
\end{equation}
where
\begin{equation}\label{eq:Bdef}
    B^{R(rm)}_k \equiv \frac{
        || ~ [T - T^{(rm)}] \ket{y^{R(rm)}_k} ||^2
    }{
        \braket{y^{R(rm)}_k | y^{R(rm)}_k}
    } ,
    \qquad
    B^{L(rm)}_k \equiv \frac{
        || \bra{y^{L(rm)}_k}  [T - T^{(rm)}] ~ ||^2
    }{
        \braket{y^{L(rm)}_k | y^{L(rm)}_k}
    }.
\end{equation}
In this form, it is clear that the residual bounds are statements about matrix elements of $T$ and $T^{(d)}$ between Ritz vectors that can be computed in any basis, not only in the Lanczos basis.
The action of $T - T^{(m)}$ on Ritz vectors can be expressed conveniently using Krylov basis vectors as
\begin{equation}
\begin{aligned}
    [T - T^{(rm)}] \ket{y^{R(rm)}_k} 
    &= \sum_{ta} \ket{k^R_{t+1,a}} P^{R(rm)}_{ta,k} - \ket{y^{R(rm)}_k} \lambda^{(rm)}_k,
    \\
    \bra{y^{L(rm)}_k} [T - T^{(rm)}] 
    &= \sum_{ta} P^{L(rm)}_{k,ta} \bra{k^L_{t+1,a}} - \lambda^{(rm)}_k \bra{y^{L(rm)}_k},
\end{aligned}
\end{equation}
noting that $T\ket{k^R_{ta}} = \ket{k^R_{t+1,a}}$ and $\bra{k^L_{ta}} T = \bra{k^L_{t+1,a}}$.
It is then straightforward to evaluate the matrix elements appearing in $B^{R/L(rm)}_k$ as 
\begin{equation}
\begin{aligned}
    & || ~ [T - T^{(rm)}] \ket{y^{R(rm)}_k} ||^2
    \equiv \braket{y^{R(rm)}_k | [T - T^{(rm)}]^\dagger [T - T^{(rm)}] |y^{R(rm)}_k}
    \\ & \quad = \sum_{ta,sb} P^{R(rm)*}_{ta,k} \left[
        \braket{k^R_{t+1,a} | k^R_{s+1,b}}
        - \lambda^{(m)}_k \braket{k^R_{t+1,a} | k^R_{s,b}} 
        - \lambda^{(m)*}_k \braket{k^R_{t,a} | k^R_{s+1,b}} 
        + |\lambda^{(m)}_k|^2 \braket{k^R_{t,a} | k^R_{s,b}} 
    \right] P^{R(rm)}_{sb,k}
    \\ & \quad \HEq \sum_{ta,sb} P^{R(rm)*}_{ta,k} \left[
        C^{RR}_{ab}(s+t+2) 
        - (\lambda^{(m)}_k + \lambda^{(m)*}_k) C^{RR}_{ab}(s+t+1) 
        + |\lambda^{(m)}_k|^2 C^{RR}_{ab}(s+t) 
    \right] P^{R(rm)}_{sb,k}
    \\ & \quad = 
        \left[ \bm{P}^{R\dagger} \bm{H}_{RR}^{(2)} \bm{P}^R \right]_{kk}
        - 2 \mathrm{Re}[\lambda^{(m)}_k] \left[ \bm{P}^{R\dagger} \bm{H}_{RR}^{(1)} \bm{P}^R \right]_{kk}
        + |\lambda^{(m)}_k|^2 \left[ \bm{P}^{R\dagger} \bm{H}_{RR}^{(0)} \bm{P}^R \right]_{kk},
\end{aligned}
\end{equation}
where $\HEq$ indicates $T = T^\dagger$ has been formally applied to take
$
    \braket{k^R_{ta}|k^R_{sb}} = \braket{\psi^R_a | (T^\dagger)^t T^s |\psi^R_b} \HEq \braket{\psi^R_a | T^{t+s} |\psi^R_b} = C^{RR}_{ab}(t+s) .
$
Similarly,
\begin{equation}
\begin{aligned}
    & || \bra{y^{L(rm)}_k}  [T - T^{(rm)}] ~ ||^2
    \equiv \braket{y^{L(rm)}_k | [T - T^{(rm)}] [T - T^{(rm)}]^\dagger |y^{L(rm)}_k}
    \\ & \quad = \sum_{ta,sb} P^{L(rm)}_{k,ta} \left[
        \braket{k^L_{t+1,a} | k^L_{s+1,b}}
        - \lambda^{(m)}_k \braket{k^L_{t+1,a} | k^L_{s,b}} 
        - \lambda^{(m)*}_k \braket{k^L_{t,a} | k^L_{s+1,b}} 
        + |\lambda^{(m)}_k|^2 \braket{k^L_{t,a} | k^L_{s,b}} 
    \right] P^{L(rm)*}_{k,sb}
    \\ & \quad \HEq \sum_{ta,sb} P^{L(rm)}_{k,ta} \left[
        C^{LL}_{ab}(s+t+2)
        - (\lambda^{(m)}_k + \lambda^{(m)*}_k) C^{LL}_{ab}(s+t+1) 
        + |\lambda^{(m)}_k|^2 C^{LL}_{ab}(s+t) 
    \right] P^{L(rm)*}_{k,sb}
    \\ & \quad = 
        \left[ \bm{P}^{L} \bm{H}_{LL}^{(2)} \bm{P}^{L\dagger} \right]_{kk}
        - 2 \mathrm{Re}[\lambda^{(m)}_k] \left[ \bm{P}^{L} \bm{H}_{LL}^{(1)} \bm{P}^{L\dagger} \right]_{kk}
        + |\lambda^{(m)}_k|^2 \left[ \bm{P}^{L} \bm{H}_{LL}^{(0)} \bm{P}^{L\dagger} \right]_{kk},
\end{aligned}
\end{equation}
using $\braket{k^L_{ta}|k^L_{sb}} \HEq \braket{\psi^L_a | T^{t+s} | \psi^L_b} = C^{LL}_{ab}(t+s)$.
This shows how the numerators of $B^{R/L(rm)}_k$ in \cref{eq:Bdef} can be computed directly from the Ritz coefficients and correlator matrices.
The norms appearing in the denominators of $B^{R/L(rm)}_k$ can be computed similarly using
\begin{equation}
    \braket{y^{R(rm)}_k | y^{R(rm)}_k}
    = \sum_{ta,sb} P^{R(rm)*}_{ta,k} \braket{k^R_{ta} | k^R_{sb}} P^{R(rm)}_{sb,k}
    \HEq \sum_{ta,sb} P^{R(rm)*}_{ta,k} C^{RR}_{ab}(t+s) P^{R(rm)}_{sb,k},
\end{equation}
and
\begin{equation}
    \braket{y^{L(rm)}_k | y^{L(rm)}_k}
    = \sum_{ta,sb} P^{L(rm)}_{k,ta} \braket{k^L_{ta} | k^L_{sb}} P^{L(rm)*}_{k,sb}
    \HEq \sum_{ta,sb} P^{L(rm)}_{k,ta} C^{LL}_{ab}(t+s) P^{L(rm)*}_{k,sb}.
\end{equation}

\end{widetext}

\section{Linear Prediction Matrices, Krylov Vectors, and Residual Bounds}
\label{app:gevp-Gammas-and-Bs}

The residual bound can also be computed directly from the Krylov-basis quantities involved in the matrix pencil $(\bm{M}, \bm{G})$  by using the Ritz coefficients. 
The denominator is straightforward to expand in the Krylov basis using the Ritz coefficients
\begin{equation}
\begin{aligned}
    \braket{y^{R(rm)}_k | y^{R(rm)}_k} &= [\bm{P}^{R(rm)\dag} \bm{G}^{RR} \bm{P}^{R(rm)}]_{kk}, \\
    \braket{y^{L(rm)}_k | y^{L(rm)}_k} &= [\bm{P}^{L(rm)} \bm{G}^{LL}  \bm{P}^{L(rm)\dag}]_{kk} ,
\end{aligned}
\end{equation}
where $G^{LL}$ and $G^{RR}$ are Hankel matrices constructed out of only the left (resp. right) Krylov vectors:
\begin{equation}
    \begin{aligned}
        G^{LL}_{ta,sb} &= \braket{k_{ta}^L | k_{sb}^L} \HEq C^{LL}_{ab}(s + t), \\
        G^{RR}_{ta,sb} &= \braket{k_{ta}^R | k_{sb}^R} \HEq C^{RR}_{ab}(s + t).
    \end{aligned}
\end{equation}
Here $C^{LL}$ and $C^{RR}$ are the diagonal correlator matrices defined in \cref{eq:corr-LL-RR-def}.

For the residual-bound numerator, the only difficulty is in determining the effect of $T - T^{(rm)}$ on an element of the Krylov basis. 
For $t < m - 1$, it is clear that $T^{(rm)} \ket{k^{R/L}_{ta}} = T \ket{k^{R/L}_{ta}} \HEq \ket{k^{R/L}_{t+1,a}}$, 
so the only difficulty is computing $T^{(rm)} \ket{k_{m-1,a}^{R/L}} \HEq \mathcal{P}^{(rm)} \ket{k_{ma}^{R/L}}$ in the Krylov basis. To calculate the projection, note that the orthogonal projection $\Pm{rm}$ preserves inner products with elements of the Krylov space, which implies
\begin{equation}
\label{eq:ortho-proj-def}
\begin{aligned}
    \braket{k_{tc}^L | \mathcal{P}^{L(rm)} | k_{ma}^L}
    &= \braket{k_{tc}^L | k_{ma}^L} = G^{LL}_{tc,ma}, \\
    \braket{k_{tc}^R | \mathcal{P}^{R(rm)} | k_{ma}^R}
    &= \braket{k_{tc}^R | k_{ma}^R} = G^{RR}_{tc,ma}.
\end{aligned}
\end{equation}
\Cref{eq:ortho-proj-def} can be expanded in the Krylov basis by defining coefficients $\Gamma^{(a)}_{tb}$ for $t \in \{0, \dots, m - 1\}$ such that 
\begin{equation}
\begin{aligned}
\mathcal{P}^{L(rm)} \ket{k_{m,a}^L} &= \sum_{t=0}^{m-1} \sum_{b} \Gamma^{L(a)}_{tb} \ket{k_{tb}^L}, \\
\mathcal{P}^{R(rm)} \ket{k_{m,a}^R} &= \sum_{t=0}^{m-1} \sum_{b} \Gamma^{R(a)}_{tb} \ket{k_{tb}^R},
\end{aligned}
\end{equation}
in which case \cref{eq:ortho-proj-def} becomes a linear equation which is solved by
\begin{equation}
\begin{aligned}
    \Gamma^{L}_{sba} &= \sum_{tc} (G^{LL})^{-1}_{sb,tc} G^{LL}_{tc,ma}, \\
    \Gamma^{R}_{sba} &= \sum_{tc} (G^{RR})^{-1}_{sb,tc} G^{RR}_{tc,ma}.
\end{aligned}
\end{equation}
These correspond precisely to the linear prediction matrices appearing in the companion matrices defined in \cref{eq:companion-block-right-def}.
This gives an alternative derivation of the fact that the companion matrix is the realization of $T^{(rm)}$ acting on Krylov space.

Once the $\bm{\Gamma}_s^{R/L}$ has been computed, the matrix elements of $T - T^{(rm)}$ in the Krylov basis is then given by
\begin{equation}
    (T - T^{(rm)}) \ket{k_{ta}^{R/L}}
    = \begin{cases}
        \sum_{s,b}  \ket{k_{sb}^{R/L}} \tilde{\Gamma}^{R/L}_{sba} & t = m \\
        0 & t < m
    \end{cases},
\end{equation}
where $\tilde{\Gamma}^{R/L}_{sba}$ is an augmented version of $\Gamma^{R/L}_{tba}$ given by
\begin{equation}
    \tilde{\Gamma}^{R/L}_{tba} = \begin{cases}
        \Gamma^{R/L}_{tba} & t < m \\
        -\delta_{ab} & t = m
    \end{cases}.
\end{equation}
Using this notation, the residual bounds can then be computed via
\begin{equation}\label{eq:resid-bound-krylov}
\begin{aligned}
    B^{R(rm)}_k &= 
    \frac{\sum_{s,t}  \vec{P}^{R(rm)\dagger}_{mk} \tilde{\bm{\Gamma}}^{R\dagger}_{t} \bm{G}^{RR}_{ts} \tilde{\bm{\Gamma}}^{R}_{s} \vec{P}_{mk}^{R(m)} }{ \sum_{s,t} \vec{P}^{L(rm)\dagger}_{tk} \bm{G}^{RR}_{ts} \vec{P}_{sk}^{R(m)} }, \\
    B^{L(rm)}_k &= 
    \frac{\sum_{s,t}  \vec{P}^{L(rm)T}_{km} \tilde{\bm{\Gamma}}^{L\dagger}_{t} \bm{G}^{LL}_{ts} \tilde{\bm{\Gamma}}^{L}_{s} \vec{P}_{km}^{L(m)*} }{ \sum_{s,t} \vec{P}^{L(rm)T}_{kt} \bm{G}^{LL}_{ts} \vec{P}_{ks}^{L(m)*} }.
\end{aligned}
\end{equation}
If $\ket{y_k^{(rm)R/L}}$ is an element of the Hermitian subspace and the $P^{R(rm)}_{ta,k}$ / $P^{L(rm)}_{k,ta}$ are normalized as described in \cref{sec:gevp}, then all quantities marked with $L$ and $R$ here are equivalent, and the denominator will always be 1, corresponding to automatic unit normalization of the Ritz vectors (see \cref{eq:ritz-norm-LR}).

The form of the residual bound in \cref{eq:resid-bound-krylov} also reveals some additional geometric structure present in the bound. In particular, both the left and right bounds contain a subexpression of the form
\begin{equation}
\begin{aligned}
    r_{ab}^{R/L}
    &= \braket{k_{ma}^{R/L} | (T - T^{(m)})^\dag (T - T^{(m)}) | k_{mb}^{R/L} } \\
    &= \sum_{ts} [ \tilde{\bm{\Gamma}}^{R/L \dagger}_{t} \bm{G}^{RR/LL}_{ts} \tilde{\bm{\Gamma}}^{R/L}_{s} ]_{ab}
\end{aligned}
\end{equation}
which is independent of the Ritz vector index $k$.
Geometrically, this factor can be interpreted as measuring how well the next set of block Krylov vectors $\{\ket{k^{R/L}_{ma}}\}_{a=1}^r$ can be reconstructed given the first $rm$ Krylov vectors. For the scalar case, this is even more apparent, since $\sin^{-1} r_{ab}$ measures the angle between $\ket{k^{R/L}_m}$ and the subspace spanned by $\{\ket{k^{R/L}_0}, \dots , \ket{k^{R/L}_{m-1}}\}$.

\section{Feynman-Hellmann}
\label{app:feynhell}

In recent years, approaches based on summing three-point correlators over the operator insertion time, like the summation~\cite{Maiani:1987by,Dong:1997xr,Capitani:2012gj,deDivitiis:2012vs}, compound propagator~\cite{Savage:2016kon,Tiburzi:2017iux}, and Feynman-Hellmann~\cite{Bouchard:2016heu,Can:2020sxc,Batelaan:2023jqp} methods, have been widely used to simplify correlator computations and analysis of excited-state effects.
One might thus consider how Prony-Ritz methods can be applied to this kind of data. Unfortunately, as shown below, we find that the methods discussed in this work are inapplicable in the forward case where the matrix elements of individual states can be isolated. In the off-forward case, Prony-Ritz methods are applicable, but individual matrix elements cannot be isolated in principle due to non-trivial mixing of quantities among the full tower of states.

In the region $0 \leq \tau \leq t$, ignoring thermal effects, a generic off-forward three-point correlator has spectral expansion
\begin{equation}
    C^\mathrm{3pt}(t,\tau) = \sum_{mn} Z^{\prime *}_n J_{nm} Z_m (\lambda'_n)^{t-\tau} \lambda_m^\tau
\end{equation}
where $\lambda_m = \exp[-E_m]$, $\lambda'_n = \exp[-E'_n]$, and unprimed/primed quantities are for the initial/final state spectrum.
The typical objective is to extract the $J_{nm}$, and especially $J_{00}$.
Compound propagator and Feynman-Hellman methods naturally compute the summed correlator $\sum_\tau C^\mathrm{3pt}(t,\tau)$, whose remaining $t$ dependence is analyzed to extract the $J_{nm}$.
Analyses of such data must consider contamination by contributions from $\tau$ outside the physical region $0 \leq \tau \leq t$; for simplicity, we ignore these contributions and consider the sum only over the physical region.
Similarly, we ignore the additional complication of contact terms which arise when $t=\tau$ and/or $\tau=0$; including them does not change the conclusions below.

Consider the sum over just the $t,\tau$-dependent part of $C^\mathrm{3pt}(t,\tau)$ in abstract, $(\lambda')^{t-\tau} \lambda^\tau$.
Two cases arise.
First, when $\lambda = \lambda'$, it evaluates simply to 
\begin{equation}
    \sum_{\tau=0}^t \lambda^{t-\tau} \lambda^\tau = (t+1) \lambda^t ~ .
\end{equation}
Second, when $\lambda \neq \lambda'$, it is a finite geometric series, which can be evaluated as
\begin{equation}\label{eq:fh-geo-sum}
    \sum_{\tau=0}^{t} (\lambda')^{t-\tau} \lambda^\tau
    = \frac{ (\lambda')^{t+1} - \lambda^{t+1} }{ \lambda' - \lambda } ~ .
\end{equation}

Prony-Ritz methods are not applicable to summed forward three-point functions where $\lambda'_n = \lambda_n$ and $Z'_n = Z_n$.
To see this, note
\begin{equation}
    \sum_{\tau=0}^t C^\mathrm{3pt}(t,\tau)
    = \sum_n |Z_n|^2 J_{nn} (t+1) \lambda_n^t + \sum_{m \neq n} (\cdots) ~ ,
\end{equation}
i.e., that the $n = m$ terms result in contributions linear in $t$.
If $|Z_n|^2 J_{nn}$ can be extracted from analysis of the time dependence, then $|Z_n|^2$ can be obtained from analysis of the corresponding two-point function and $J_{nn}$ isolated.
However, Prony-Ritz methods necessarily produce an exponential decomposition of the data.
There is no limit where a single exponential contribution $\sim \lambda^t$ becomes linear.
A Prony-Ritz decomposition of the data will then necessarily include several exponential contributions with fine-tuned near-cancellations to mock up the linear time dependence.
The amplitudes of these terms do not (in any obviously useful way) isolate or approximate the desired quantity, $|Z_n|^2 J_{nn}$.
A similar issue arises in applying RR to summed ratios of three- and two-point functions as arise in the summation method; for summed ratios, linear terms arise also in the off-forward case.

The issue of linear contributions does not arise in the off-forward case where $\lambda'_n \neq \lambda_m$ for any $n,m$ (up to numerical coincidences, which we ignore).
However, a separate issue prevents clean isolation of the desired information.
Summing over $\tau$ gives
\begin{widetext}
\begin{equation}
\begin{aligned}
    \sum_{\tau=0}^t C^\mathrm{3pt}(t, \tau)
    &= \sum_{mn} Z^{\prime *}_n J_{nm} Z_m 
        \frac{ (\lambda'_n)^{t+1} - \lambda_m^{t+1} }{ \lambda'_n - \lambda_m }
    \\
    &= \sum_{mn} \frac{ Z^{\prime *}_n J_{nm} Z_m }{ \lambda'_n - \lambda_m } \, (\lambda'_n)^{t+1}
     ~ - ~ \sum_{mn} \frac{ Z^{\prime *}_n J_{nm} Z_m }{ \lambda'_n - \lambda_m } \, \lambda_m^{t+1}
    \\
    &= \sum_{n} Z^{\prime *}_n  \left[ \sum_m \frac{  J_{nm} Z_m }{ \lambda'_n - \lambda_m } \right] \, (\lambda'_n)^{t+1}
     ~ - ~ \sum_{m} \left[ \sum_n\frac{ Z^{\prime *}_n J_{nm}  }{ \lambda'_n - \lambda_m } \right] Z_m \, \lambda_m^{t+1}
    \\
    &\equiv \sum_{n} Z^{\prime *}_n  (\mathcal{JZ})_n \, (\lambda'_n)^{t+1}
     ~ - ~ \sum_{m} (\mathcal{ZJ})^*_m Z_m \, \lambda_m^{t+1} ~ .
\end{aligned}
\end{equation}
\end{widetext}
In the first equality, we use \cref{eq:fh-geo-sum}.
In the second equality, we recognize that the contribution for each state splits apart into two terms with different time dependence, such that the spectrum visible in the summed correlator is the union of the initial- and final-state spectra.
The ground-state matrix element is present in two different terms, i.e.~those for the ground states of each of the initial- and final-state spectra.
In the third equality, in each term, we group terms for the state index $m$ or $n$ which is not associated with the time dependence.
In the fourth equality, we define the symbols
\begin{equation}
    \begin{aligned}
        (\mathcal{JZ})_n &\equiv \sum_m \frac{  J_{nm} Z_m }{ \lambda'_n - \lambda_m } 
        \\
        (\mathcal{ZJ})^*_m &\equiv \sum_n \frac{ Z^{\prime *}_n J_{nm}  }{ \lambda'_n - \lambda_m } ~ .
    \end{aligned}
\end{equation}
Analysis of the time dependence of the summed correlator can only ever in principle extract the overall overlap for each separate time dependence, i.e.~can constrain $Z^{\prime *}_n (\mathcal{JZ})_n$ and $(\mathcal{ZJ})^*_m Z_m$.
Analysis of the initial- and final-state two-points gives $Z^{\prime *}_n$ and $Z_m$, which allows isolating $(\mathcal{JZ})_n$ and $(\mathcal{ZJ})^*_m$.
However, each of these quantities involves a sum over the full spectral tower, mixing together the overlaps and matrix elements of infinitely many states.
Similar challenges to obtaining off-forward matrix elements with summation-type methods are discussed in Ref.~\cite{Tiburzi:2017iux}.
Mixing into the ground-state matrix element is only suppressed if the overlaps with excited states are small, or by the factor $1/(\lambda'_n - \lambda_m)$.

\section{GEVP Facts}
\label{app:gevp}

Consider the generalized eigenvalue problem
\begin{equation}
    \bm{M} \vec{v} = \lambda \bm{G} \vec{v}
\end{equation}
where $\bm{M}$ and $\bm{G}$ are both Hermitian, $n \times n$ matrices. If $\bm{G}$ is further assumed positive-definite, the GEVP is called \emph{definite}, and standard results show that the generalized eigenvectors $\vec{v}_i$ form a basis, and furthermore can be taken to satisfy the orthogonality condition
\begin{equation}
    \vec{v}_i^\dag \bm{G} \vec{v}_j = \delta_{ij}
\end{equation}
In general, however, the GEVPs considered in this work will not be definite in the presence of noise. In that case, the set of possibilities is considerably more complicated; for instance, if $\bm{M}$ and $\bm{G}$ share a common null space, then the GEVP is called \emph{ill-disposed} and can become arbitrarily ill-conditioned~\cite{Parlett,stewart1973subspace,bai2000templates}.
Ill-disposed eigenproblems are of some concern here, since Rayleigh-Ritz approximations are often approximately ill-disposed due to the near-degenerate nature of Krylov space~\cite{STEWART1978193}.
In practice, however, the ill-conditioning applies eigenvalue by eigenvalue, meaning that some subset of eigenvalues will become arbitrarily ill-conditioned, while others remain relatively stable under perturbations (see Refs.~\cite{stewart1990matrix,stewart1979pertubation,stewart1973subspace} for more information on the perturbation theory for GEVPs).
This gives further merit to the idea that some eigenvalues are entirely spurious, arising solely out of numeric instability and not from any physical phenomenon.

Regardless of the conditioning of the problem, the generalized eigenvalues are roots of the polynomial
\begin{equation}
    p(\lambda) = \det (\bm{M} - \lambda \bm{G}).
\end{equation}
Unless the problem is ill-disposed (in which case $p(\lambda) = 0$ identically), there will be exactly $n$ eigenvalues counted with multiplicity. 
For practical purposes, we take the additional assumptions that $\bm{G}$ is invertible, and the GEVP is non-degenerate, meaning that there are $n$ distinct roots, which also guarantees the existence of a basis of generalized eigenvectors. Note that the set on which this condition fails is a set of measure 0, and hence nondegeneracy will be generically satisfied by noisy data.

By the assumption that $\bm{M}$ and $\bm{G}$ are Hermitian, it is straightforward to see that
$p(\lambda)^* = p(\lambda^*)$, which implies eigenvalues are either real or come in complex-conjugate pairs. Furthermore,
if $(v_1, \lambda_1)$ and $(v_2, \lambda_2)$ are distinct eigenpairs, then the computation
\begin{equation}
    \lambda_1^* v_1^\dag \bm{G} v_2
    = v_1^\dag \bm{M} v_2
    = \lambda_2 v_1^\dag \bm{G} v_2
\end{equation}
shows that either $v_1^\dag \bm{G} v_2 = 0$ or $\lambda_1 = \lambda_2^*$.
Combined with the assumed nondegeneracy of the eigenproblem, this implies that any right eigenvector with generalized eigenvalue $\lambda$ is also a left eigenvector with eigenvalue $\lambda^*$.
Furthermore, this also implies that different eigenvectors are $\bm{G}$-orthogonal, meaning that if the right and left eigenvectors are arranged into a matrices $\bm{P}^R$ and $\bm{P^L}$ then 
\begin{equation}
\label{eq:gevp-partial-norm}
    [(\bm{P}^L)^\dag \bm{G} \bm{P^R}]_{ij} = c_i \delta_{ij},
\end{equation}
with arbitrary constants $c_i \in \mathbb{C}$. By assumption, $\bm{G}, \bm{P}^L$, and $\bm{P}^R$ are invertible, so the RHS of \cref{eq:gevp-partial-norm} must be invertible as well, which implies $c_i \neq 0$. Applying an appropriate rescaling to either the left or right eigenvectors, this proves it is always possible to enforce the normalization condition
\begin{equation}
    [(\bm{P}^L)^\dag \bm{G} \bm{P^R}]_{ij} = \delta_{ij}
\end{equation}
which corresponds to the biorthogonality of the Ritz vectors, as discussed in \cref{sec:gevp}.

\section{Computational Complexity}
Given the many equivalent formulations of RR, it is natural to consider what asymptotic complexity is achievable for each step of the algorithm.
The application of the Lanczos recursions is $O(m^2)$ as highlighted in Ref.~\cite{Ostmeyer:2024qgu}.
It is further possible to diagonalize Hermitian tridiagonal matrices in approximately $O(m^{2.3})$ time or even $O(m \ln^p m)$ using divide-and-conquer algorithms~\cite{Demmel:1997}; however these algorithms do not apply to the non-Hermitian tridiagonal matrices that arise in oblique Lanczos.
In the case of non-Hermitian $T^{(d)}$ relevant to noisy LQCD correlator analysis, it is not obvious whether the Ritz values can be computed using the oblique (block) Lanczos algorithm with fewer than the $O(m^3)$ operations required to diagonalize $T^{(d)}$ using e.g.~the QR algorithm.

In general, most methods for computing the Ritz values and other associated quantities can be reduced to a small number of matrix operations, resulting in $O(m^3)$ time complexity in practice.\footnote{Here ``matrix operation'' refers to either a $m \times m$ matrix multiplication, or any $m \times m$ matrix decomposition, all of which inherit the same asymptotic time complexity, either $O(m^3)$ naively, $O(m^{\log_2 7})$ using Strassen-type algorithms~\cite{strassen1969gaussian}, or lower using Coppersmith-Winograd-type algorithms~\cite{coppersmith1987matrix}. For simplicity, we assume all matrix operations take $O(m^3)$ time, which is true for most algorithms used in practice.} For a lower bound on the asymptotic complexity, computing $m$ Ritz values clearly requires $\Omega(m)$ work, and it seems unlikely that any method can run faster than $O(m^2)$ as that would require avoiding any direct construction of any matrix.

Given the prevalence of matrix operations in all the methods presented, it would seem plausible to conjecture that the Ritz values cannot be computed any faster than a matrix operation, effectively $O(m^3)$. This, however, is false, and in fact it is possible to compute both the Ritz values and the $Z$-factors in $O(m^2)$ time for the scalar case using the companion matrix formalism of Prony's method. 

To see how $O(m^2)$ scaling can be achieved using the companion matrix, first note that the linear system \cref{eq:scalar-companion-hankel-action} is a Hankel system, which is equivalent to a Toeplitz system and hence is amenable to Levinson recursion, which solves for one column of the output in $O(m^2)$ time~\cite{levinson1946,Golub:2013}. Only one column of the companion matrix is nontrivial (see \cref{eq:companion-scalar-def}), so this gives access to the companion matrix, from which the Ritz values can be computed in $O(m^2)$ using the methods of Ref.~\cite{aurentz2015fast}.
Once the Ritz values are determined, computing the $Z$-factors can be accomplished via \cref{eq:scalar-vandermonde-system}, which requires solving a Vandermonde system, which can be accomplished in $O(m^2)$ time~\cite{Golub:2013}. Lastly, individual columns of the right Ritz coefficients can be computed in $O(m^2)$ time using the fact that the eigenvectors of a companion matrix are an inverse Vandermonde matrix~\cite{Golub:2013}, which allows the computation of the matrix elements for an individual state in $O(m^2)$ time via \cref{eq:matrix-elements-PrPl}, though computing all $m$ possible matrix elements is a full matrix operation, and hence would take $O(m^3)$ time.

Note that although the methods above do have better asymptotic complexity, this does not necessarily imply that they are better to use in practice. More specialized algorithms tend to to be less numerically stable, potentially necessitating custom implementations in high precision even in cases where the Hankel GEVP can be computed in double precision. Extending these methods to the block case is also unclear; Levinson recursion can be generalized to block matrices (see Ref.~\cite{akaike1973block}), but computing the $Z$-factors would also necessitate generalizing the Vandermonde structure in an appropriate manner. At the scale of current lattice calculations, using $O(m^3)$ matrix operations seems sufficient for practical purposes, but future investigation is needed to determine if and when these asymptotically efficient algorithms are more useful in practice.

\bibliography{Lanczos_bib}

\end{document}